\documentclass[twocolumn, aps, prd,superscriptaddress]{revtex4-1}
\usepackage{graphicx}
\usepackage{amsmath}
\usepackage{color}
\usepackage[colorlinks=true,citecolor=blue,urlcolor=blue,breaklinks]{hyperref}

\begin{document}
\title{Linear perturbations of  low angular momentum accretion flow in the Kerr metric and the corresponding emergent gravity phenomena}%

\author{Md Arif Shaikh}%
\email[]{arifshaikh@hri.res.in}
\affiliation{Harish-Chandra Research Institute, HBNI, Chhatnag Road, Jhunsi, Allahabad 211019, India}
\author{Tapas Kumar Das}
\email{tapas@hri.res.in}
\affiliation{Harish-Chandra Research Institute, HBNI, Chhatnag Road, Jhunsi, Allahabad 211019, India}
\affiliation{Physics and applied mathematics unit, Indian Statistical Institute, Kolkata 700108, India}
\date{\today}

\begin{abstract}
\noindent For certain geometric configuration of matter falling onto a rotating black hole, we develop a novel linear perturbation analysis scheme to perform the stability analysis of stationary integral accretion solutions corresponding to the steady state low angular momentum, inviscid, barotropic, irrotational, general relativistic accretion of hydrodynamic fluid. We demonstrate that such steady states remain stable under linear perturbation, and hence the stationary solutions are reliable to probe the black hole spacetime using the accretion phenomena.We report that a relativistic acoustic geometry emerges out as the consequence of such stability analysis procedure. We study various properties of that sonic geometry in detail. We construct the causal structures to establish the one to one correspondences of the sonic points with the acoustic black hole horizons, and the shock location with an acoustic white hole horizon. The influence of the spin of the rotating black holes on the emergence of such acoustic spacetime has been discussed.  
\end{abstract}

\maketitle

\section{Introduction}
Understanding the accretion process is important to study the observational signature of astrophysical black holes (\cite{Frank1985,Mineshige2008}). One studies the dynamical and the radiative properties of black hole accretion to construct the characteristic black hole spectra, such spectra is analyzed observationally to probe the spacetime at the close proximity of the horizon of astrophysical black holes. Because of the inner boundary conditions posed by the presence of the event horizon, black hole accretion is necessarily transonic (\cite{Liang1980}) except for the possible cases of wind fed accretion of supersonic stellar winds (\cite{Morris1996}). 

For low angular momentum accretion, flow can manifest multi-transonicity, i.e., one may observe the transition from the subsonic to supersonic flow at more than one place during the course of motion of the matter falling towards the horizon, originating out from the infinity (from a reasonably large distance). Accretion solutions passing through more than one sonic points may be connected through a discontinuous shock wave (\cite{Liang1980,Abramowicz1981,Muchotrzeb1982,Muchotrzeb1983,Fukue1983,Fukue1987,Lu1985,Lu1986,Muchotrzeb1986,Abramowicz1989,Abramowicz1990,CHAKRABARTI_physics_Reports,Kafatos1994,Yang1995,Pariev1996,Peitz1997,Caditz1998,Das2002,Das2003,Barai2004,Fukue2004,abraham2006,Das2007,Okuda2004,Okuda2007,Das2012,Sukova2015JOP,Sukova2015MNRAS,Sukova2017}).Study of the aforementioned shocked multi-transonic accretion is, usually, accomplished for steady state matter flow, and the stationary integral accretion solutions are considered to probe the shock formation phenomena as well as to construct the corresponding black hole spectra (\cite{CHAKRABARTI_physics_Reports} and the references therein), although full time-dependent numerical simulation works are also performed to understand several characteristic features of the black hole accretion in general. For low angular momentum stationary integral accretion solutions, it is usually assumed that the flow in inviscid, barotropic and irrotational.

It is relevant to note in this aspect that nonsteady features (time variability) and various kind of local as well as global fluctuations may be present for large-scale astrophysical fluid flows, which may jeopardize the steady state, and the stationary integral flow solutions may not be used  to construct the black hole spectra in such circumstances. To ensure that one can use the stationary integral flow solutions to study the black hole accretion in a reliable way, it is thus imperative to establish that the steady state accretion model under consideration remains stable under perturbation.

In our present work, we would like to investigate the effect of the linear perturbation on stationary accretion solutions obtained for steady state general relativistic accretion onto rotating astrophysical black holes, i.e., accretion flow studied in the Kerr metric. For low angular momentum inviscid accretion, conical wedge-shaped flow is ideal to simulate the geometric configuration of matter accreting onto the black hole. It is to be mentioned here that by `low' angular momentum, we essentially mean accretion flow with sub-Keplerian angular momentum distribution.  For such flow, the stable Keplerian orbit may not form and hence it is not necessary to introduce the viscous dissipation to make the stable circular orbit collapse and to let the matter accrete onto the black hole. The inviscid assumption is considered to be justified for such flow profile. It is believed that such flow structure is a common feature for accretion onto the supermassive black hole at our Galactic centre (see \cite{Moscibrodzka2006} and references therein).  Such sub-Keplerian weakly rotating flows may be observed  in various astrophysical systems, for  detached binary systems fed by accretion from OB stellar winds ( \cite{Illarionov1975a,Liang1984}), for instance. Also for  semi-detached low-mass non-magnetic binaries (\cite{Bisikalo1998}),
and for super-massive black holes fed by accretion from slowly rotating central stellar clusters (\cite{Illarionov1988,Ho1999} and references therein) such flows are common. Even for a standard Keplerian accretion disc, turbulence may produce such low angular momentum flow (see, e.g., \cite{Igumenshchev1999} and references therein). In supersonic astrophysical flows, perturbation of various kinds may produce discontinuities.  The type of low angular momentum flow which will be discussed in the present work is somewhat different from thick accretion disc models in the sense that considerable amount of radial advective velocity is included as the initial boundary condition for our flow model. Such advection may be a consequence of high-velocity stellar wind fed accretion. Such advective accretion flows in the Kerr metric with complete general relativistic treatment for shock formation in conical flow has not been treated in the literature before. 

The equations governing such flow will be derived from the first principle and the stationary integral flow solutions corresponding to the steady state will be obtained. It will be demonstrated that for certain values of initial conditions governing the flow, such integral solutions may pass through two sonic points, and flows passing through two sonic points will be connected through a stationary shock wave. Such stationary integral solutions will then be linear perturbed to demonstrate that the perturbation does not diverge, which ensures that the stationary integral solutions are reliable because the corresponding steady state remains stable under linear perturbation. While performing the aforementioned procedure of linear stability analysis, one observes, as will be demonstrated in the consequent sections, the linear perturbation (the corresponding `sound wave') propagates within the accreting fluid with a certain speed of propagation. It is also observed that the propagation of such linear acoustic perturbation can be described using a particular kind of acoustic spacetime (conformal to a certain form of Black hole metric), that spacetime is further described by a metric. One can write down the specific form of such acoustic metric embedded within the background stationary accreting flow.

Such findings lead to very interesting consequences. In the field of analogue gravity phenomena, it has been suggested that a black hole like spacetime can be generated within a transonic fluid by linear perturbing such flow. The propagation of linear perturbation within such fluid can be described using a spacetime metric, conformal to the Painleve-Gullstrand-Lemaitre (\cite{Painleve,Gullstrand,Lemaitre}) presentation of Schwarzschild metric (\cite{Unruh,Visser1998,Bilic1999,Barcelo,Novello-visser,Unruh-Schutzhold,Analogue-gravity-phenomenology}). The acoustic metric can possess corresponding acoustic horizons, depending on certain criteria, such horizons may be of the black hole or white hole types (\cite{Barcelo2004}). Acoustic black holes are formed where the background fluid makes a transition from the subsonic to the supersonic state and acoustic white holes are formed where the background fluid may make a transition from the supersonic state to the subsonic state. 

Since the Hawking effect, as well as its counterpart in analogue gravity, is a kinematical effect, one can define the acoustic surface gravity in a more general way, i.e., acoustic surface gravity can be evaluated on any kind of acoustic horizon. Following such approach, it has also been stated in the literature that depending on certain initial condition, there is a theoretical probability that a classical analogue system can have infinitely large (or, at least, extremely large) value of acoustic surface gravity (\cite{Liberati2000}).

Aforementioned information leads to the conclusion that not only an accreting black hole system can be considered as a classical analogue model, but the acoustic geometry is a natural consequence of the process of linear perturbation of the stationary integral accretion solutions of steady, inviscid, barotropic, irrotational flow of matter onto astrophysical black holes. In the present work, we precisely demonstrate that. We discuss that the emergent gravity phenomena are the natural consequence of the linear stability analysis of steady-state accretion, and hence make the crucial connection between two apparently disjoint fields of research, namely, the astrophysical accretion process and the emergent gravity phenomenon. We make a formal correspondence between the sonic surfaces in accretion astrophysics with the acoustic black hole horizons in analogue gravity, between the discontinuous stationary shock in multi-transonic black hole accretion with the acoustic white hole in the analogue spacetime, through the construction of the corresponding causal structure at and around the sonic points and the shock locations, respectively.

In the subsequent section, we show that the acoustic surface gravity corresponding to the accreting black hole system can be calculated in terms of the gradient of background steady state fluid velocity as well as that of the sound speed, both evaluated at the acoustic horizon. At sonic points, the transition from the subsonic to the supersonic state is continuous, and hence such gradients, as well as the acoustic surface gravity have a finite value. At the location of the formation of the stationary shock, there is a discontinuous transition from supersonic state to the subsonic state corresponding to the stationary integral solutions. The gradient of the fluid, as well as the sound velocity, diverge at the shock location. As a result, the corresponding value of the acoustic surface gravity evaluated at the shock location (acoustic white hole) is infinite. This is an important finding since it manifests that the theoretical results obtained in \cite{Liberati2000} are actually relevant for a realistic physical system as well.

In accretion astrophysics, it is usually believed that the majority of the black holes contain non-zero spin angular momentum, i.e., most of the astrophysical black holes are of Kerr type (\cite{Miller2009,Kato2010,Ziolkowski2010,Tchekhovskoy2010,
	Daly2011,Buliga2011,Reynolds2012,McClintock2011,Martínez-Sansigre2011,Dauser2010,Nixon2011,McKinney2012,
	McKinney2013,Brenneman,Rolando2013,Sesana,Fabian2014,
	Healy,Jiang,Nemmen}). 
It is thus important to understand how the black hole spin (the Kerr parameter) influences the overall features of the emergent gravity phenomena as observed while linear perturbing the background solutions in the Kerr metric. 

In the present work, we will demonstrate the black hole spin dependence of the corresponding analogue gravity phenomena. In this way, we also try to understand how the properties of the background black hole metric influence the characteristic feature of the sonic metric embedded within the accreting fluids.

In what follows, we demonstrate how one obtains the governing equations corresponding to the low angular momentum, inviscid, polytropic, irrotational, axially symmetric, non self-gravitating general relativistic accretion of hydrodynamic fluid onto a rotating black hole in the background Kerr metric. We shall set $G=c=M_{\rm BH}=1$ where $G$ is the universal gravitational constant, $c$ is the velocity of light and $M_{\rm BH}$ is the mass of the black hole. The radial distance will be scaled by $G M_{\rm BH}/c^2$ and any velocity will be scaled by $c$. We shall use the negative-time-positive-space metric convention.

\section{Basic equations governing the flow}
We consider the following metric for a stationary rotating spacetime
\begin{equation}\label{background_metirc}
ds^2=-g_{tt}dt^2+g_{rr}dr^2+g_{\theta\theta}d\theta^2+2g_{\phi t}d\phi dt+g_{\phi\phi}d\phi^2,
\end{equation}
where the metric elements are function of $r,\theta$ and $\phi$. The metric elements in the Boyer-Lindquist coordinates are given by (\cite{Boyer-Lindquist})
\begin{equation}\label{back_metric_elements}
\begin{aligned}
& g_{tt}=\left(1-\frac{2}{\mu r}\right),\quad
g_{rr}=\frac{\mu r^2}{\Delta},\quad
g_{\theta\theta}=\mu r^2,\\
& g_{\phi t}=g_{t\phi}=-\frac{2a\sin^2\theta}{\mu r},\quad
g_{\phi\phi}=\frac{\Sigma}{\mu r^2}\sin^2\theta
\end{aligned}
\end{equation}
where
\begin{equation}\label{A-B-C}
\begin{aligned}
& \mu =1+\frac{a^2}{r^2}\cos^2\theta,\quad \Delta=r^2-2r+a^2,\\
&\Sigma=(r^2+a^2)^2-a^2\Delta\sin^2\theta.
\end{aligned}
\end{equation}
The event horizon of the Kerr black hole is located at $ r_+ = 1+\sqrt{1-a^2}  $.

We assume the hydrodynamic fluid accreting onto the Kerr black hole to be perfect, irrotational, and is described by adiabatic equation of state. The energy momentum tensor for such fluid is given by
\begin{equation}\label{Energy_momentum_tensor}
 T^{\mu\nu}=(p+\varepsilon)v^\mu v^\nu+pg^{\mu\nu}
 \end{equation}
 where $p$ and $\varepsilon$ are the pressure and the energy density of the fluid, respectively. $v^\mu$ is the four-velocity of the fluid which satisfies the normalization condition $v^\mu v_\mu=-1$. The adiabatic equation of state is given by the relation $p = k\rho^\gamma$ where $\rho$ is the rest-mass energy density and $\gamma = c_p/c_v$ is the adiabatic index ($ c_p $ and $ c_v $ are specific heats at constant pressure and at constant volume, respectively ). The total energy density $ \varepsilon $ is the sum of the rest-mass energy density and the internal energy density (due to the thermal energy), i.e., $ \varepsilon = \rho + \varepsilon_{\rm thermal} $. The continuity equation which ensures the conservation of mass is given by
\begin{equation}\label{Continuity_eq}
\nabla_\mu(\rho v^\mu)=0
\end{equation}
where the covariant divergence is defined as $\nabla_\mu v^\nu = \partial_\mu v^\nu + \Gamma^\nu_{\mu\lambda}v^\lambda$ with the Christoffel symbols $\Gamma^{\nu}_{\mu\lambda} = \frac{1}{2}g^{\nu \sigma}[\partial_\lambda g_{\sigma \mu} + \partial_\mu g_{\sigma \lambda} - \partial_\sigma g_{\mu\lambda}]$. The energy-momentum conservation equation is given by
\begin{equation}\label{mom_consv_eq}
\nabla_\mu T^{\mu\nu}=0.
\end{equation}
Substitution of Eq. (\ref{Energy_momentum_tensor}) in Eq. (\ref{mom_consv_eq}) provides
 the general relativistic Euler equation for barotropic ideal fluid as
\begin{equation}\label{Euler_eq}
(p+\varepsilon)v^\mu \nabla_\mu v^\nu+(g^{\mu\nu}+v^\mu v^\nu) \nabla_\mu p=0.
\end{equation}
The specific enthalpy of the flow is defined as $ h=(p+\varepsilon)/\rho$. We assume the flow to be isentropic, i.e., the specific entropy of the flow $s/\rho$ is constant, where $ s $ is the entropy density. Therefore, for an isentropic flow, the following thermodynical identity, where $T$ is the temperature of the fluid,
\begin{equation}
dh = T d(\frac{s}{\rho})+\frac{1}{\rho}dp
\end{equation}
gives $dp = \rho dh$ which when used in $ h=(p+\varepsilon)/\rho$ also gives $ d\varepsilon = \rho dh $.  Thus the adiabatic sound speed is given by 
\begin{equation}
c_s^2 =\left. \frac{\partial p}{\partial \varepsilon}\right|_{\frac{s}{\rho} = {\rm constant}} = \frac{\rho}{h}\frac{\partial h}{\partial \rho}.
\end{equation}
The relativistic Euler equation for isentropic flow can thus be written as
\begin{equation}\label{Euler_eq_for_isothermal}
v^\mu \nabla_\mu v^\nu+\frac{c_s^2}{\rho}(v^\mu v^\nu+g^{\mu\nu})\partial_\mu \rho=0.
\end{equation}
For general relativistic irrotational isentropic fluid, the irrotationality condition is given by (\cite{Bilic1999})
\begin{equation}\label{irrotationality_condition}
\partial_\mu(h v_\nu)-\partial_\nu(h v_\mu)=0.
\end{equation}

\section{Accretion flow geometry}
We consider an axially symmetric accretion flow in the Kerr background. The flow is assumed to be symmetric about the equatorial plane. The four velocity components are written as ($v^t,v^r,v^\theta,v^\phi$). We assume that the velocity component along the vertical direction is negligible compared to the radial component $v^r$, i.e., $v^\theta\ll v^r$. Also due the axial symmetry $\partial_\phi$ term in the continuity equation given by Eq. (\ref{Continuity_eq}) would vanish. Thus the continuity equation for such flow can be written as
\begin{equation}\label{cont_eq_wo_av}
\partial_t(\rho v^t\sqrt{-g})+\partial_r(\rho v^r \sqrt{-g}) = 0
\end{equation}
where $g$ is the determinant of the metric $g_{\mu\nu}$. For Kerr metric, $g=-\sin^2\theta r^4 \mu^2$.
The accretion flow variables, i.e., velocity components and the density are in general functions of $ t,r,\theta $ coordinates. However, assuming that the flow thickness is small compared to the radial size of the accretion disc, one can do an averaging of any flow variable $ f(t,r,\theta) $ along the $ \theta $ direction by the following approximation (\cite{gammie1998} )
\begin{equation}\label{vertical-averaging}
\int f(t,r,\theta)d\theta \approx H_\theta f(t,r,\theta=\frac{\pi}{2})
\end{equation}
where $H_\theta$ is the characteristic angular scale of the flow. Such an averaging is very common in accretion disc literature and is usually known as vertical averaging of the flow. Thus the continuity equation for vertically averaged axially symmetric accretion can be written as (\cite{Deepika2017,Shaikh2017})
\begin{equation}\label{continuity_eq_equatorial}
\partial_t(\rho v^t \sqrt{-\tilde{g}}H_\theta)+\partial_r(\rho v^r \sqrt{-\tilde{g}} H_\theta)=0
\end{equation}
where $\tilde{g}$ is the value of $g$ on the equatorial plane, i.e., $\tilde{g}=-r^4$. The advantage of vertical averaging is that all the variables are defined by their values measured on the equatorial plane and any information about the geometry of the flow along the vertical direction is contained in the term $H_\theta$. $H_\theta$ is a function of the local flow thickness $H(r)$. The angular scale $ H_\theta $ of the flow thickness, i.e., the angle made by the flow thickness at the center of the black hole at any radial distances $ r $ from the center of the black hole along the equatorial plane is given by $ H_{\theta} = {H(r)}/{r} $, assuming the the flow thickness to be small at all $ r $.  

There are mainly three models of flow geometry in the literature. The simplest one is called the constant height model (CH). For such flow geometry, the flow thickness remains constant for all $ r $, i.e., for such flow $ H(r) = {\rm constant} $ or $ H_\theta \propto {1}/{r} $. In the second model, the flow geometry is considered to be quasi-spherical or wedge-shaped conical. Such flow is known as conical flow (CF). For conical flow geometry, the angular scale $ H_\theta $ remains constant at all $ r $. In other words, the flow thickness is proportional to the radial distance or $ H(r)\propto r $. The third model is the most complicated one. In this model, the flow is considered to be in hydrostatic equilibrium in the vertical direction and is known as vertical equilibrium model (VE). Further details on such classification are available in \cite{Bilic2014}. In VE model, the local flow thickness will depend on the flow variables also.  The general relativistic calculation by \cite{Abramowicz1996ap} for stationary case gives the flow thickness for VE model through the following relation
\begin{equation}
-\frac{2p}{\rho}+\frac{H_{\rm VE}^2(r)}{r^4}(v_\phi^2-a^2(v_t^2-1)) = 0
\end{equation}
As evident from the above relation, the flow thickness in VE model is a complicated function of the flow variables $ p,\rho,v_\phi,v_t $ as well as the black hole spin $ a $.

In the present work, we consider the conical flow model where the accretion flow is assumed to maintain a wedge-shaped conical geometry. As mentioned earlier, in such flow the local flow thickness is proportional to the radial distance measured along the equatorial plane, i.e., $\frac{H}{r} = {\rm constant}$ or $H_\theta$ being the characteristic angular scale of local flow is constant for such conical flow geometry. Thus $H_\theta$ does not depend on the accretion flow variables like velocity or density. Therefore linear perturbation of these quantities (discussed in the next section) will have no effect on it. For simplicity, therefore, we will write $H_\theta$ simply as $H_0$. The same is true for the CH model also. However, due to the complicated dependence of $ H(r) $ on the flow variables in the VE model, the flow thickness will also be perturbed when the flow variables are perturbed. This will make the analysis too complicated to be presented here and may be reported elsewhere. Therefore as mentioned earlier, we do not consider CH and VE models and work only with the CF model. From now on all the equations will be derived by assuming that the flow variables are vertically averaged and their values are computed on the equatorial plane.

\section{Linear perturbation analysis and the acoustic geometry}
The scheme of the linear perturbation analysis would be the following: We shall write the accretion variables, e.g., four velocity components and density about their stationary background values upto first order in perturbation. These expressions are then used in the basic governing equations such as the continuity equation, normalization condition and the irrotationality condition. Keeping only the terms that are linear in the perturbed quantities gives equations relating different perturbed quantities upto first order in perturbations. Further manipulations of these equations gives a wave equation which describes the propagation of the perturbation of the mass accretion rate which is defined later in this section. Such wave equation mimics the wave equation for a massless scalar field in curved spacetime. Finally comparing theses two wave equations one obtains the acoustic metric.   

Below we derive some useful relations using the irrotationality condition (Eq. (\ref{irrotationality_condition})), the normalization condition $v^\mu v_\mu=-1$ and the axial symmetry which will be later used to derive the wave equation for linear perturbation.
From irrotaionality condition given by Eq. (\ref{irrotationality_condition}) with $\mu=t$ and $\nu=\phi$ and with axial symmetry we have
\begin{equation}\label{irrotaionality_t_phi}
\partial_t(h v_\phi)=0,
\end{equation}
again with $\mu=r$ and $\nu=\phi$ and the axial symmetry the irrotationality condition gives
\begin{equation}\label{irrotationality_r_phi}
\partial_r(h v_\phi)=0.
\end{equation}
So we get that $h v_\phi$ is a constant of the motion. Eq. (\ref{irrotaionality_t_phi}) gives
\begin{equation}\label{del_t_v_phi}
\partial_t v_\phi=-\frac{v_\phi c_s^2}{\rho}\partial_t \rho.
\end{equation}
Substituting  $v_\phi=g_{\phi\phi}v^\phi+g_{\phi t}v^t$ in the above equation provides
\begin{equation}\label{del_t_v_up_phi}
\partial_t v^\phi=-\frac{g_{\phi t}}{g_{\phi\phi}}\partial_t v^t-\frac{v_\phi c_s^2}{g_{\phi\phi}\rho}\partial_t \rho.
\end{equation}
The normalization condition $v^\mu v_\mu=-1$ provides
\begin{equation}\label{Normalization_condition}
g_{tt}(v^t)^2=1+g_{rr}(v^r)^2+g_{\phi\phi}(v^\phi)^2+2g_{\phi t}v^\phi v^t
\end{equation}
which after differentiating with respect to $t$ gives
\begin{equation}\label{del_t_v_up_t_1}
\partial_t v^t=\alpha_1\partial_t v^r+\alpha_2\partial_t v^\phi
\end{equation}
where $\alpha_1=-\frac{v_r}{v_t}$, $\alpha_2=-\frac{v_\phi}{v_t}$ and $v_t = -g_{tt}v^t+g_{\phi t}v^\phi$. Substituting $\partial_t v^\phi$ in Eq. (\ref{del_t_v_up_t_1}) using Eq. (\ref{del_t_v_up_phi}) gives \begin{equation}\label{del_t_v_up_t_2}
\partial_t v^t=\left( \frac{-\alpha_2v_\phi c_s^2/(\rho g_{\phi\phi})}{1+\alpha_2 g_{\phi t}/g_{\phi\phi}}\right)\partial_t \rho+\left(\frac{\alpha_1}{1+\alpha_2 g_{\phi t}/g_{\phi\phi}} \right)\partial_t v^r
\end{equation}
We perturb the velocities and density around their steady background values as following
\begin{equation}\label{pert_v}
v^\mu(r,t)=v^\mu_0(r)+v^\mu_1(r,t) 
\end{equation}
\begin{equation}\label{pert_rho}
\rho(r,t)=\rho_0(r)+\rho_1(r,t)
\end{equation}
 where $ \mu = t,r,\phi $ and the subscript ``0" denotes the stationary background part and the subscript ``1" denotes the linear perturbations. Using Eq. (\ref{pert_v})-(\ref{pert_rho}) in Eq. (\ref{del_t_v_up_t_2}) and retaining only the terms of first order in perturbed quantities we obtain

\begin{equation}\label{del_t_pert_v_up_t}
\partial_t v_1^t=\eta_1\partial_t \rho_1+\eta_2\partial_t v^r_1
\end{equation}
where
\begin{equation}\label{eta_1_eta_2_and_Lambda}
\begin{aligned}
&\eta_1=-\frac{c_{s0}^2}{\Lambda v^t_0\rho_0}[\Lambda (v^t_0)^2-1-g_{rr}(v^r_0)^2],\quad \eta_2=\frac{g_{rr}v^r_0}{\Lambda v^t_0}\\
& {\rm and}\quad \Lambda=g_{tt}+\frac{g_{\phi t}^2}{g_{\phi\phi}}
\end{aligned}
\end{equation}
\subsection{Linear perturbation of mass accretion rate}\label{sec:mass}
For stationary background flow the $\partial_t$ part of the equation of continuity, i.e., Eq. (\ref{continuity_eq_equatorial}) vanishes and integration over spatial coordinate provides 
$\sqrt{-\tilde{g}}H_0 \rho_0 v^r_0 ={\rm constant}.$
Multiplying the quantity $\sqrt{-\tilde{g}}H_0\rho_0 v^r_0$ by the azimuthal component of volume element $d\phi$ and integrating the final expression gives the mass accretion rate,  $\Psi_0=\tilde{\Omega}\sqrt{-\tilde{g}}H_0\rho_0v^r_0$. $\Psi_0$ gives the rate of infall of mass through a particular surface. $\tilde{\Omega}$ arises due to the integral over $\phi$ and is just a geometrical factor and therefore can we can redefine the mass accretion rate by setting it to unity without any loss of generality. Thus we define
\begin{equation}
\Psi_0\equiv \sqrt{-\tilde{g}}H_0\rho_0v^r_0.
\end{equation}
Now let us define a quantity $\Psi\equiv\sqrt{-\tilde{g}}H \rho(r,t) v^r(r,t)$ which has the stationary value equal to $\Psi_0$. Using the perturbed quantities given by Eq. (\ref{pert_v}) and (\ref{pert_rho})  we have
\begin{equation}\label{pert_mass_accretion_rate}
\Psi(r,t)=\Psi_0+\Psi_1(r,t),
\end{equation}
where
\begin{equation}\label{psi_1}
\Psi_1(r,t)=\sqrt{-\tilde{g}}H_0(\rho_0 v^r_1+v^r_0 \rho_1).
\end{equation}
Using Eq. (\ref{pert_v})-(\ref{del_t_pert_v_up_t}) and (\ref{pert_mass_accretion_rate}) in the continuity Eq. (\ref{continuity_eq_equatorial}) and  differentiating Eq. (\ref{psi_1}) with respect to $t$ gives, respectively
\begin{equation}\label{del_r_psi_1}
\rho_0\eta_2\partial_t v^r_1+(v^t_0+\rho_0\eta_1)\partial_t \rho_1=-\dfrac{1}{\sqrt{-\tilde{g}}H_0}\partial_r \Psi_1,
\end{equation}
and
\begin{equation}\label{del_t_psi_1}
\rho_0\partial_t v^r_1+ v^r_0\partial_t \rho_1=\dfrac{1}{\sqrt{-\tilde{g}}H_0}\partial_t\Psi_1.
\end{equation}
In deriving Eq. (\ref{del_r_psi_1}) we have used Eq. (\ref{del_t_pert_v_up_t}). With these two equations given by Eq. (\ref{del_r_psi_1}) and (\ref{del_t_psi_1}) we can write $\partial_t v^r_1$ and $\partial_t \rho_1$ solely in terms of derivatives of $\Psi_1$ as
\begin{equation}\label{del_t_rho_1_and_v_1}
\begin{aligned}
& \partial_t v^r_1=\frac{1}{\sqrt{-\tilde{g}}H_0\rho_0 \tilde{\Lambda}}[-(v^t_0+\rho_0\eta_1)\partial_t\Psi_1-v^r_0\partial_r\Psi_1]\\
& \partial_t \rho_1=\frac{1}{\sqrt{-\tilde{g}}H_0\rho_0 \tilde{\Lambda}}[\rho_0\eta_2\partial_t\Psi_1+\rho_0\partial_r\Psi_1]
\end{aligned}
\end{equation}
where $\tilde{\Lambda}$ is given by
\begin{equation}\label{Lambda_tilde}
\tilde{\Lambda}=\frac{g_{rr}(v^r_0)^2}{\Lambda v^t_0}-v^t_0+\frac{c_{s0}^2}{\Lambda v^t_0}[\Lambda (v^t_0)^2-1-g_{rr}(v^r_0)^2].
\end{equation}
Now let us go back to the irrotationality condition given by the Eq. (\ref{irrotationality_condition}). Using $\mu=t$ and $\nu=r$  gives the following equation
\begin{equation}\label{w_mass_1}
 \partial_t(hg_{rr}v^r)-\partial_r(hv_t)=0
 \end{equation}
For stationary flow this provides $\xi_0 = -h_0v_{t0}={\rm constant}$ which is the specific energy of the system. 
We substitute the density and velocities in Eq. (\ref{w_mass_1}) using Eq. (\ref{pert_v}), (\ref{pert_rho})  and 
\begin{equation}
 \label{pert_v_t_lower}
 v_t(r,t)=v_{t0}(r)+v_{t1}(r,t).
 \end{equation}
Keeping only the terms that are linear in the perturbed quantities and differentiating with respect to time $t$ gives   
 \begin{equation}\label{w_mass_2}
 \begin{aligned}
 &\partial_t\left(h_0g_{rr}\partial_t v^r_1 \right)+\partial_t\left( \frac{h_0g_{rr}c_{s0}^2 v^r_0}{\rho_0}\partial_t \rho_1\right)\\
 &-\partial_r\left( h_0\partial_t v_{t1}\right)-\partial_r\left( \frac{h_0 v_{t0}c_{s0}^2}{\rho_0}\partial_t \rho_1\right)=0
 \end{aligned}
 \end{equation}
 We can also write
 \begin{equation}\label{del_t_pert_v_lower_t}
 \partial_t v_{t1}=\tilde{\eta}_1\partial_t \rho_1+\tilde{\eta}_2 \partial_t v^r_1
 \end{equation}
 with
 \begin{equation}\label{eta_2_tilde_and_eta_2_tilde}
 \tilde{\eta}_1=-\left(\Lambda \eta_1+\frac{g_{\phi t}v_{\phi 0}c_{s0}^2}{g_{\phi\phi}\rho_0} \right),\quad\tilde{\eta}_2=-\Lambda \eta_2.
 \end{equation}
Using Eq. (\ref{del_t_pert_v_lower_t}) in the Eq. (\ref{w_mass_2}) and dividing the resultant equation by $h_0 v_{t0}$ provides
\begin{equation}\label{w_mass_3}
\begin{aligned}
& \partial_t\left(\frac{g_{rr}}{v_{t0}}\partial_t v^r_1 \right)+\partial_t\left( \frac{g_{rr}c_{s0}^2 v^r_0}{\rho_0 v_{t0}}\partial_t \rho_1\right)\\
&-\partial_r\left( \frac{\tilde{\eta}_2}{v_{t0}}\partial_t v^r_1\right)
-\partial_r\left( (\frac{\tilde{\eta_1}}{v_{t0}}+\frac{c_{s0}^2}{\rho_0})\partial_t \rho_1\right)=0
 \end{aligned}
\end{equation}
where we have used that $h_0v_{t0}={\rm constant}$. Finally substituting $\partial_t v^r_1$ and $\partial_t \rho_1$ in Eq. (\ref{w_mass_3}) using Eq. (\ref{del_t_rho_1_and_v_1}) we get
\begin{equation}\label{w_mass_final}
\begin{aligned}
&\partial_t\left[ k(r)\left(-g^{tt}+(v^t_0)^2(1-\frac{1}{c_{s0}^2}) \right)\right]\\
&+\partial_t\left[ k(r)\left(v^r_0v_0^t(1-\frac{1}{c_{s0}^2}) \right)\right] \\
&+\partial_r \left[ k(r)\left(v^r_0v_0^t(1-\frac{1}{c_{s0}^2}) \right)\right]\\
&+\partial_r \left[ k(r)\left( g^{rr}
+(v^r_0)^2(1-\frac{1}{c_{s0}^2})\right)\right]=0
\end{aligned}
\end{equation}
where
\begin{equation}\label{conformal_mass}
k(r)=\frac{g_{rr}v^r_0c_{s0}^2}{v^t_0 v_{t0}\tilde{\Lambda}} \quad {\rm and}\quad g^{tt}=\frac{1}{\Lambda}=\frac{1}{g_{tt}+g_{\phi t}^2/g_{\phi\phi}}
\end{equation}
Eq. (\ref{w_mass_final}) can be written as $\partial_\mu (f^{\mu\nu}\partial_\nu \Psi_1)=0$ where $f^{\mu\nu}$ is given by the symmetric matrix

\begin{equation}\label{f_mass}
\begin{aligned}
&f^{\mu\nu}=\frac{g_{rr}v^r_0c_{s0}^2}{v^t_0 v_{t0}\tilde{\Lambda}}\\
&\times\left[\begin{array}{cc}
-g^{tt}+(v^t_0)^2(1-\frac{1}{c_{s0}^2}) & v^r_0v_0^t(1-\frac{1}{c_{s0}^2})\\
v^r_0v_0^t(1-\frac{1}{c_{s0}^2}) & g^{rr}+(v^r_0)^2(1-\frac{1}{c_{s0}^2})
\end{array}\right]
\end{aligned}
\end{equation}
This is the main result of this section and will be used in the next section to obtain the acoustic metric and in Sec. \ref{sec:stability} for linear stability analysis of the stationary accretion solutions in the Kerr metric. In the Schwarzschild limit ($a=0$) we have $v_{t0}\tilde{\Lambda}=1+(1-c_{s0}^2)g_{\phi\phi}(v^\phi_0)^2$. Thus the $f^{\mu\nu}$ in Eq. (\ref{f_mass}) matches  the result obtained by  \cite{Deepika2017} in the Schwarzschild limit.
\subsection{The acoustic metric}
The linear perturbation analysis performed in the previous section provides the equation describing the propagation of the linear perturbation of the mass accretion rate $ \Psi_1(r,t) $ and is given by the following equation 
\begin{eqnarray}
\partial_\mu(f^{\mu\nu}\partial_\nu \Psi_1)=0
\end{eqnarray}
where $ \mu,\nu$ each runs over $ r,t $. This equation could be compared to the wave equation of a massless scalar field $\varphi$ propagating in a curved spacetime given by (\cite{birrell1984quantum})
\begin{equation}
\partial_\mu (\sqrt{-{g}}g^{\mu\nu}\partial_\nu \varphi)=0.
\end{equation}
Thus, comparing these two equations one obtains the acoustic metric $G^{\mu\nu}$ which is related to $ f^{\mu\nu} $ in the following way
\begin{equation}
 \sqrt{-G}G^{\mu\nu}=f^{\mu\nu},
 \end{equation} 
where $G$ is the determinant of the acoustic metric $G_{\mu\nu}$. $ f^{\mu\nu} $ could be written as $ f^{\mu\nu} = k(r)\tilde{f}^{\mu\nu} $, where $ k(r) $ is the overall multiplicative factor and $ \tilde{f}^{\mu\nu} $ is the matrix part as given in Eq. \ref{f_mass}. Thus $ G^{\mu\nu} = (k(r)/\sqrt{-G})\tilde{f}^{\mu\nu}  $ and therefore, $ G^{\mu\nu} $ is related to $ \tilde{f}^{\mu\nu} $ by a conformal factor given by $ k(r)/\sqrt{-G} $. One of our main goals of the present work is to show that the acoustic horizon are the transonic surface of the accretion flow and to demonstrate that by studying the causal structure of the acoustic spacetime. However, the location of the event horizon or the causal structure of the spacetime do not depend on the conformal factor of the spacetime metric. Thus in order to investigate these properties of the acoustic spacetime we can take $ G^{\mu\nu} $ to be the same as $ \tilde{f}^{\mu\nu} $ by ignoring the conformal factor. Thus the acoustic metric $G^{\mu\nu}$ and $G_{\mu\nu}$, apart from the conformal factor, are given by 
\begin{equation}
G^{\mu\nu}=\left[\begin{array}{cc}
-g^{tt}+(v^t_0)^2(1-\frac{1}{c_{s0}^2}) & v^r_0v_0^t(1-\frac{1}{c_{s0}^2})\\
v^r_0v_0^t(1-\frac{1}{c_{s0}^2}) & g^{rr}+(v^r_0)^2(1-\frac{1}{c_{s0}^2})
\end{array}\right]
\end{equation} 
and
\begin{equation}\label{acoustic_metric}
G_{\mu\nu}=\left[
\begin{array}{cc}
-g^{rr}-(v^r_0)^2(1-\frac{1}{c_{s0}^2}) & v^r_0v_0^t(1-\frac{1}{c_{s0}^2})\\
v^r_0v_0^t(1-\frac{1}{c_{s0}^2}) & g^{tt}-(v^t_0)^2(1-\frac{1}{c_{s0}^2}) 
\end{array} \right]
\end{equation}
\section{Location of the acoustic event horizon}
The metric corresponding to the acoustic spacetime is given by Eq. (\ref{acoustic_metric}). The metric elements of $G_{\mu\nu}$  are independent of time and thus the metric is stationary. In general relativity, the event horizon for such stationary spacetime is defined as a time like  hypersurface $r={\rm constant}$ whose normal $n_{\mu}=\partial_\mu r =\delta^r_\mu$ is null with respect to the spacetime metric. In the similar way we can define the event horizon of the acoustic spacetime as a null timelike hypersurface. Thus the location of the acoustic horizon is given by the condition (\cite{Moncrief1980,carroll2004,Poisson2004,abraham2006}
\begin{equation}
G^{\mu\nu}n_{\mu}n_\nu = G^{\mu\nu}\delta^r_\mu\delta^r_\nu = G^{rr}=0.
\end{equation}
Therefore on the event horizon we have the following condition
\begin{equation}\label{horizon:1}
c_{s0}^2=\frac{g_{rr}(v^r_0)^2}{1+g_{rr}(v^r_0)^2}.
\end{equation}

Now it is convenient to move to the co-rotating frame as defined in \cite{gammie1998}. Let $ u $ be the radial velocity of the fluid in the co-rotating frame which is referred as the `advective velocity' and $\lambda=-{v_{\phi}}/{v_{t}}$ be the specific angular momentum. For stationary flow the advective velocity and the specific angular momentum will be denoted with a subscript ``0" as earlier. In this co-rotating frame  we can write $v^r,v^t$ and $ v_t $ in terms of $ u,\lambda $ as 
\begin{equation}\label{v_in_CRF}
v^r=\frac{u}{\sqrt{g_{rr}(1-u^2)}}
\end{equation}

\begin{equation}\label{vt_0}
v^t=\sqrt{\frac{(g_{\phi\phi}+\lambda g_{\phi t})^2}{(g_{\phi\phi}+2\lambda g_{\phi t}-\lambda^2g_{tt})(g_{\phi \phi}g_{tt}+g_{\phi t}^2)(1-u^2)}}
\end{equation}
and 
\begin{equation}\label{v_t}
v_t = -\sqrt{\frac{g_{tt}g_{\phi\phi}+g_{\phi t}^2}{(g_{\phi\phi}+2\lambda g_{\phi t}-\lambda^2 g_{tt})(1-u^2)}}
\end{equation}
In co-rotating frame the Eq. \ref{horizon:1} becomes
\begin{equation}
u_0^2|_{\rm h}=c_{s0}^2|_{\rm h}.
\end{equation}
where the subscript ``h" implies that the quantity is to be evaluated at the horizon and would imply the same hereafter.
Thus we see that the acoustic horizon is located at a radius where the adevcetive velocity $ u_0 $ becomes equal to the speed of sound $ c_{s0} $ which is exactly the surface known as the transonic surface. Thus the transonic surface of the accretion flow and the acoustic horizon coincide.
\section{Causal structure of the acoustic spacetime}\label{Sec:Causal_structure}
Acoustic null geodesic corresponding to the radially traveling ($d\phi=0,d\theta=0$) acoustic phonons is given by 
$ds^2=0$. Thus 
\begin{equation}\label{drdt_pm}
(\frac{dr}{dt})_\pm\equiv b_\pm=\frac{-G_{rt}\pm \sqrt{G_{rt}^2-G_{rr}G_{tt}}}{G_{rr}}
\end{equation}
where the acoustic metric elements $G_{tt},G_{rt}=G_{tr},G_{rr}$ are given by Eq. (\ref{acoustic_metric}). These are expressed in terms of the background metric elements, the sound speed and the velocity variables $u_0(r)$ and $\lambda_0 = -v_{\phi0}/v_{t0}$ using Eq. (\ref{v_in_CRF}) and (\ref{vt_0})
\begin{equation}\label{metric_elements_in_u_lambda}
\begin{aligned}
& G_{tt} = -\frac{1}{g_{rr}(1-u_0^2)}\left(1-\frac{u_0^2}{c_{s0}^2} \right)\\
& G_{tr}=G_{rt} = \frac{u_0}{(1-u_0^2)}\left(1-\frac{1}{c_{s0}^2}\right)\times \\
&\sqrt{\frac{(g_{\phi\phi}+\lambda_0 g_{\phi t})^2}{g_{rr}(g_{\phi\phi}+2\lambda_0 g_{\phi t}-\lambda_0^2g_{tt})(g_{\phi \phi}g_{tt}+g_{\phi t}^2)}}\\
& G_{rr} = \frac{1}{g_{tt}g_{\phi\phi}+g_{\phi t}^2}\times \\ 
& \left(g_{\phi\phi}-\frac{(g_{\phi\phi}+\lambda_0 g_{\phi t})^2}{(g_{\phi\phi}+2\lambda_0 g_{\phi t}-\lambda_0^2 g_{tt})}\frac{(1-\frac{1}{c_{s0}^2})}{(1-u_0^2)} \right).
\end{aligned}
\end{equation}
$t(r)$ can be obtained as 
\begin{equation}\label{t_pm}
t(r)_\pm=t_0+\int_{r_0}^r \frac{dr}{b_\pm}.
\end{equation}
We can introduce a new set of coordinates as following
\begin{equation}\label{dudw}
dz = dt-\frac{1}{b_+}dr,\quad{\rm and},\quad dw = dt-\frac{1}{b_-}dr
\end{equation}
In terms of these new coordinates the acoustic line element can be written as
\begin{equation}
ds^2|_{\phi=\theta={\rm const}}=\mathcal{D}dzdw
\end{equation}
Where $\mathcal{D}$  is found to be equal to $G_{tt}$.

$b_\pm(r)$ is function of the stationary solution $u_0(r)$ and the sound speed $c_{s0}(r)$. Therefore we have to first obtain $u_0(r)$ and $c_{s0}(r)$ for stationary accretion flow. This is done by simultaneously numerically integrating the equations describing the gradient of the advective velocity $ du_0/dr $ and the gradient of the sound speed $ dc_{s0}/dr $ which are derived in Appendix \ref{Sec:flow-eq}. We use $4^{\rm {th}}$ order Runge-Kutta method to integrate these equations. The solutions are characterized by the parameters [$\xi_0$, $\gamma$, $\lambda_0$, $a$]. Remember that $\xi_0=-h_0v_{t0}$ is the specific energy of the flow which is a conserved quantity for the flow under consideration. Thus given a particular set of [$\xi_0$, $\gamma$, $\lambda_0$, $a$] we get $u_0(r)$ and $c_{s0}(r)$ by numerically solving the Eq. (\ref{dudr}) and (\ref{dcdr}) simultaneously and then using these solutions of $u_0(r)$ and $c_{s0}(r)$ we get $b_\pm(r)$. The integration in Eq. (\ref{t_pm}) is then performed by applying Euler method. Finally we plot $t(r)_\pm$ as function of $r$ to see the causal structure of the acoustic spacetime. 

\subsection{Mono-transonic case}
Let us first consider the case where the accretion flow is mono-transonic. For such accretion flow there exist only one transonic surface. In other words, the flow starts its journey from large radial distance subsonically, i.e., $ |u_0|<|c_{s0}| $ or $ \mathcal{M}= |u_0|/|c_{s0}|<1 $, where $ \mathcal{M} $ is the Mach number of the flow and at some certain radial distance $ r $, the advective velocity becomes equal to the speed of sound or $ \mathcal{M} = 1 $. The radius $ r $ at which $ \mathcal{M} $ becomes equal to 1 is called the transonic point. For the flow under consideration, the transonic points are the critical points of the flow where the denominator in the expression of $ du_0/dr $ becomes 0 (see Appendix \ref{Sec:flow-eq}). Thus the transonic points are given by $ r=r_{\rm crit} $ which in turn are obtained by solving the Eq. (\ref{critical_points}) for given values of the parameters $ [\xi_0,\gamma,\lambda_0,a] $. For $ r<r_{\rm crit} $ the flow is supersonic, i.e., $ \mathcal{M}>1 $ and remains supersonic all the way upto the event horizon $ r_+ $. 

\begin{figure}
	\centering
	\includegraphics[scale = 0.4]{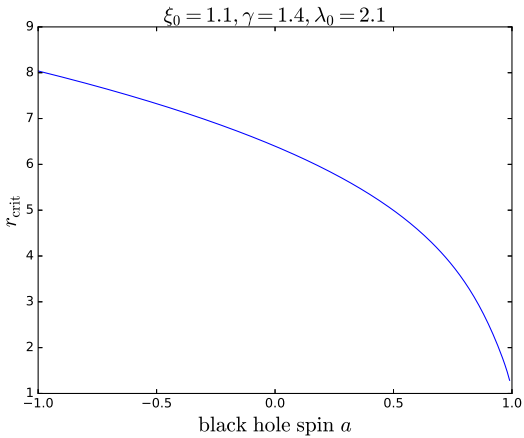}
	\caption{ The critical points $ r_{\rm crit} $ (which are transonic points of the mono-transonic accretion flow) is plotted as function of the black hole spin $ a $ for the set of values $ [\xi_0,\gamma,\lambda_0]=[1.1,1.4,2.1] $. }\label{Fig:mono-tran}
\end{figure}

We would like to choose the parameters $ [\xi_0,\gamma,\lambda_0] $ in a way such that the Eq. (\ref{critical_points}) has exactly one solution outside the event horizon (i.e., for $ r>r_+ $) for all values of $ a $ and see how the radius of the transonic surface or equivalently $ r_{\rm crit} $ varies with the black hole spin $ a $. Then with the same $ [\xi_0,\gamma,\lambda_0] $ we pick a few values of the black hole spin $ a $ and draw the causal structure of the acoustic space time and show that the location of the acoustic horizon matches $ r_{\rm crit} $ for that value of $ a $.  In Fig. \ref{Fig:mono-tran}, we plot the critical points $ r_{\rm crit} $ of mono-transonic flow as a function of the black hole spin. 

\begin{figure*}
	\centering
	\begin{tabular}{cc}
		\includegraphics[scale = 0.4]{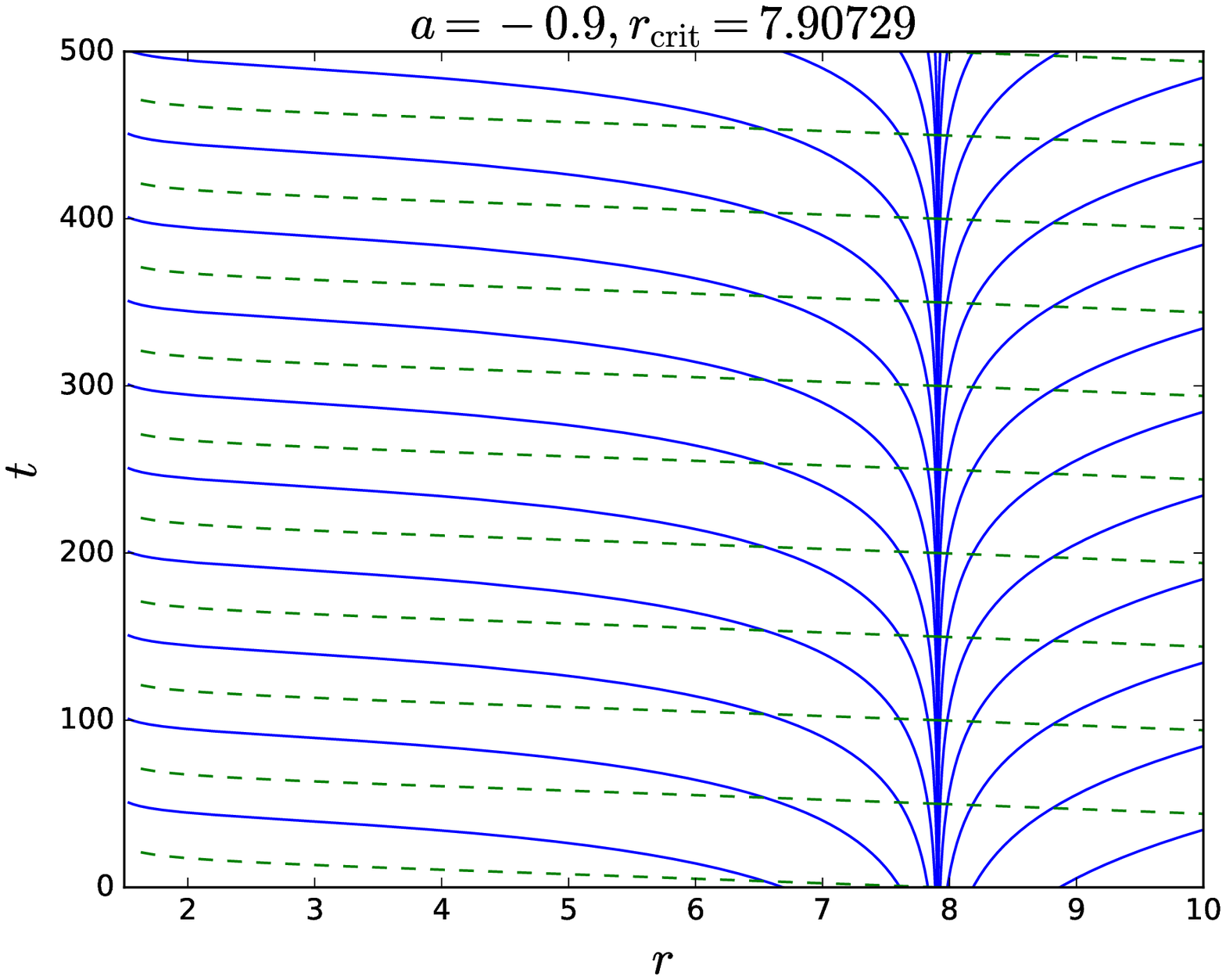} & \includegraphics[scale = 0.4]{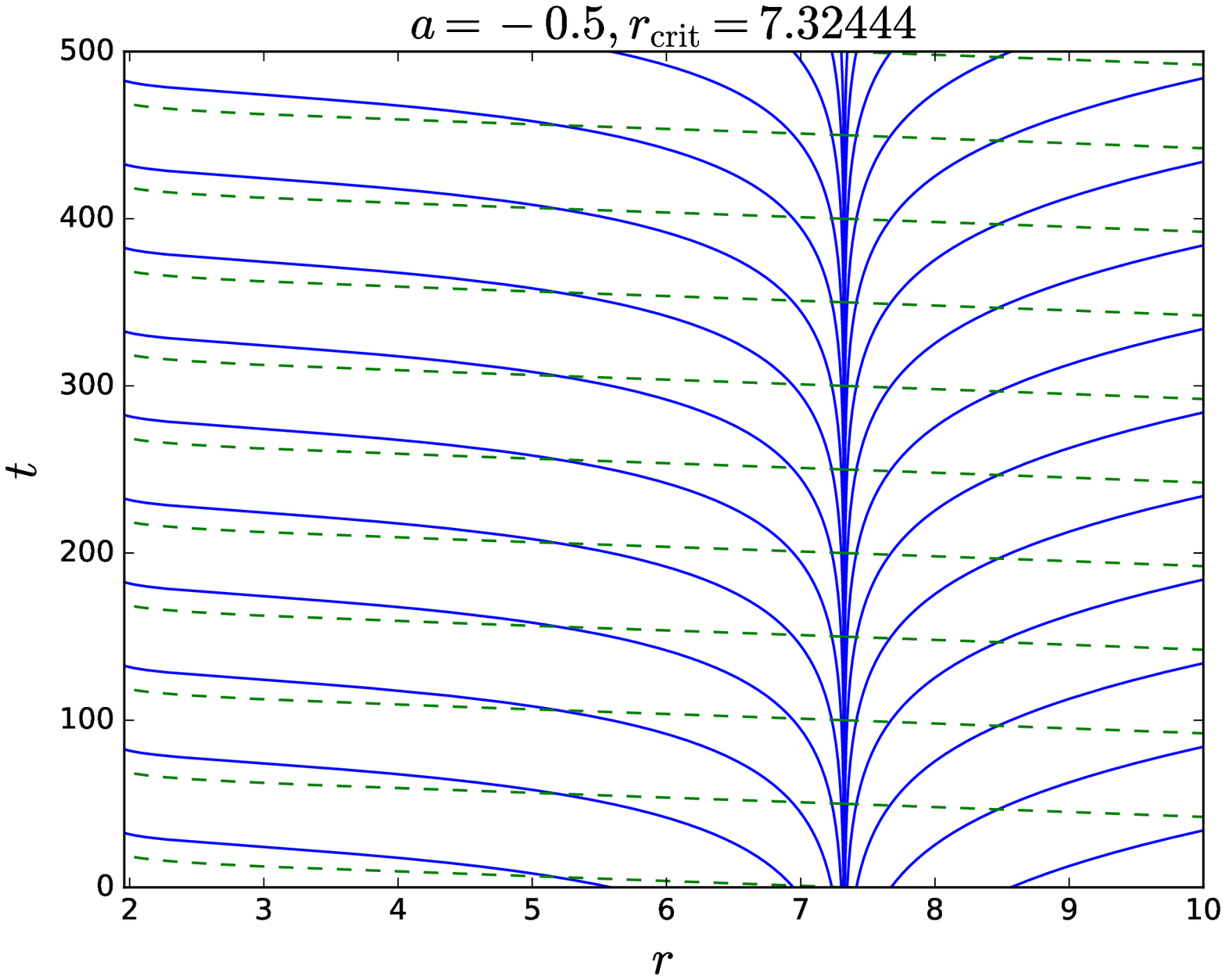}\\
		\includegraphics[scale = 0.4]{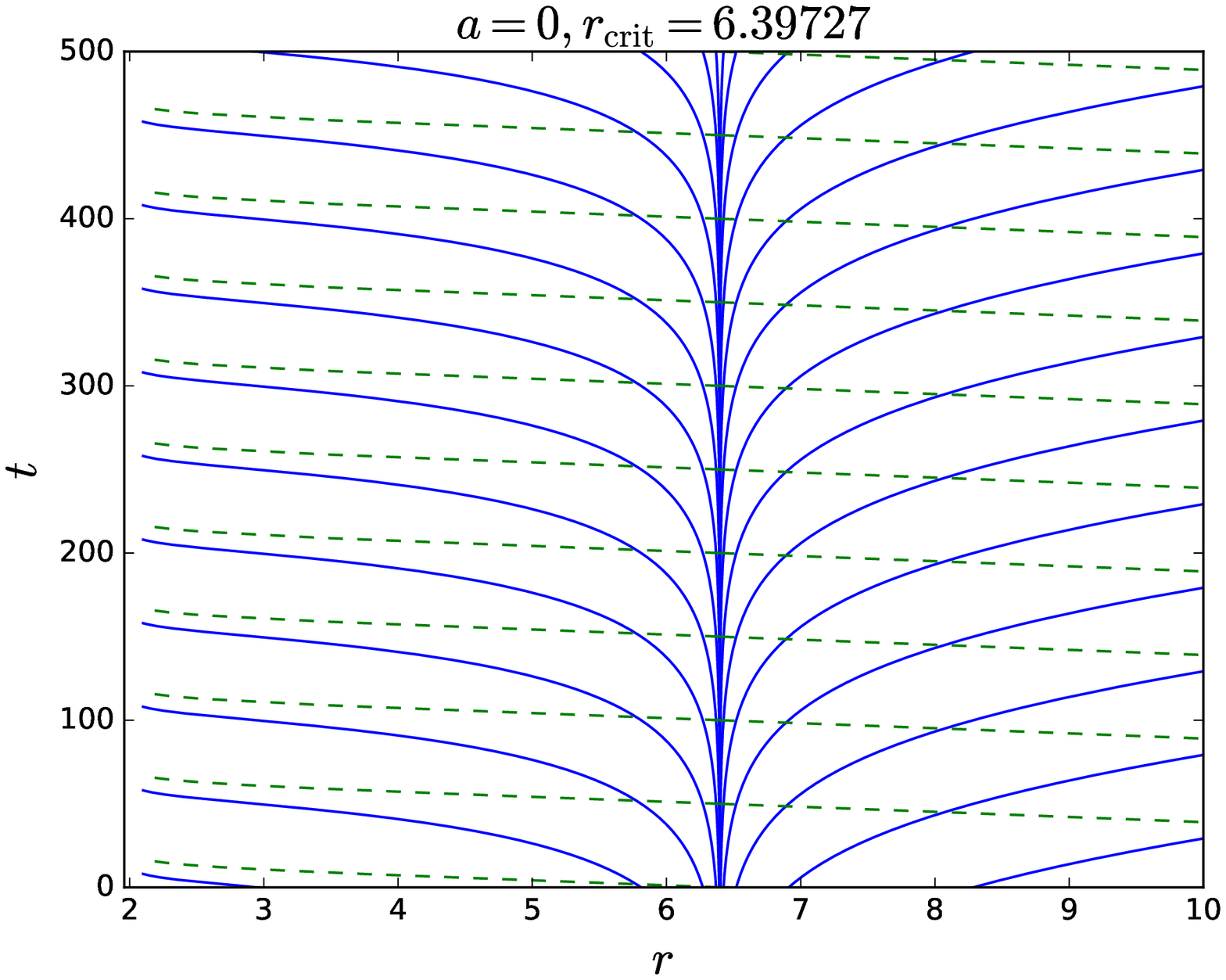} & \includegraphics[scale = 0.4]{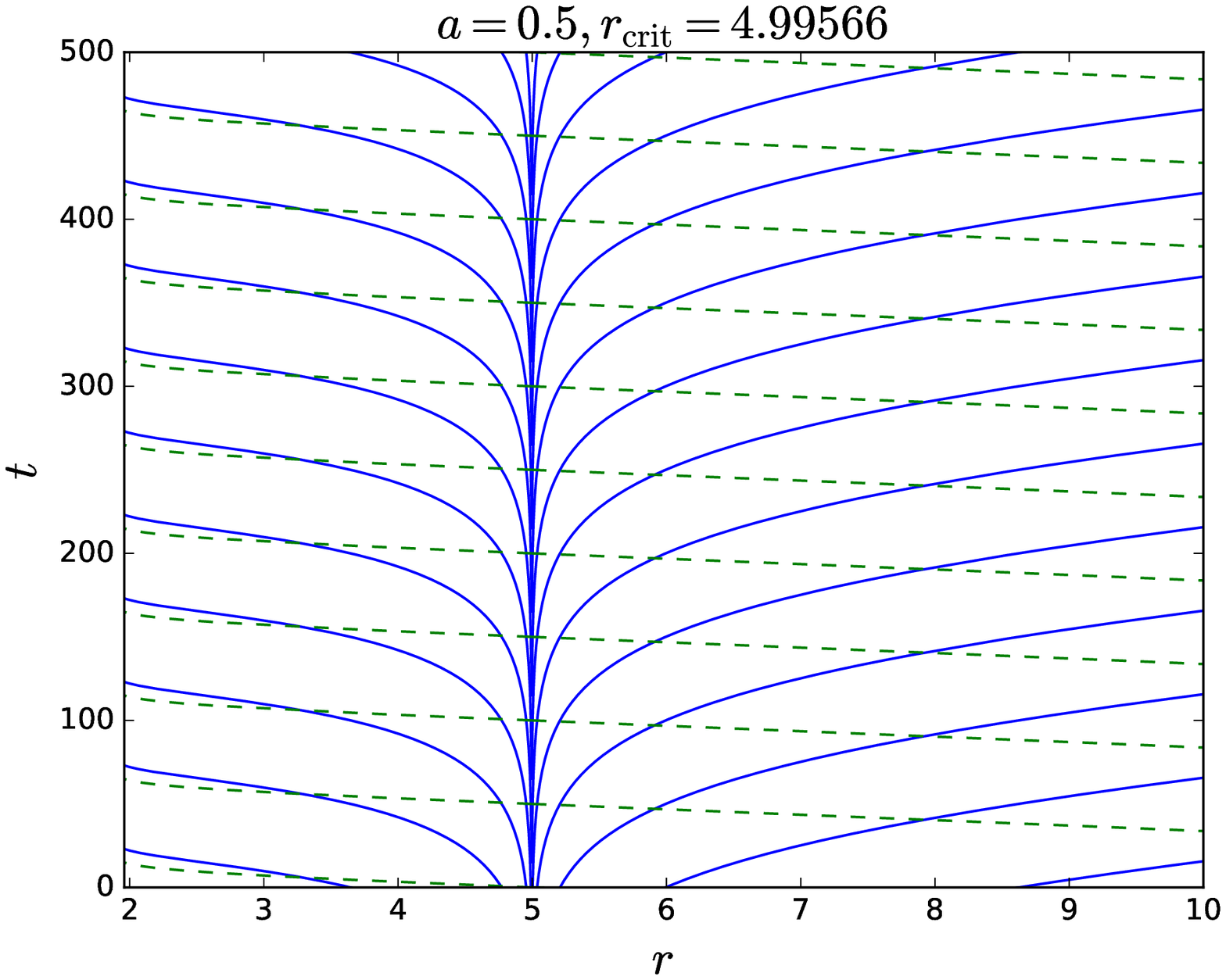}\\
		\includegraphics[scale = 0.4]{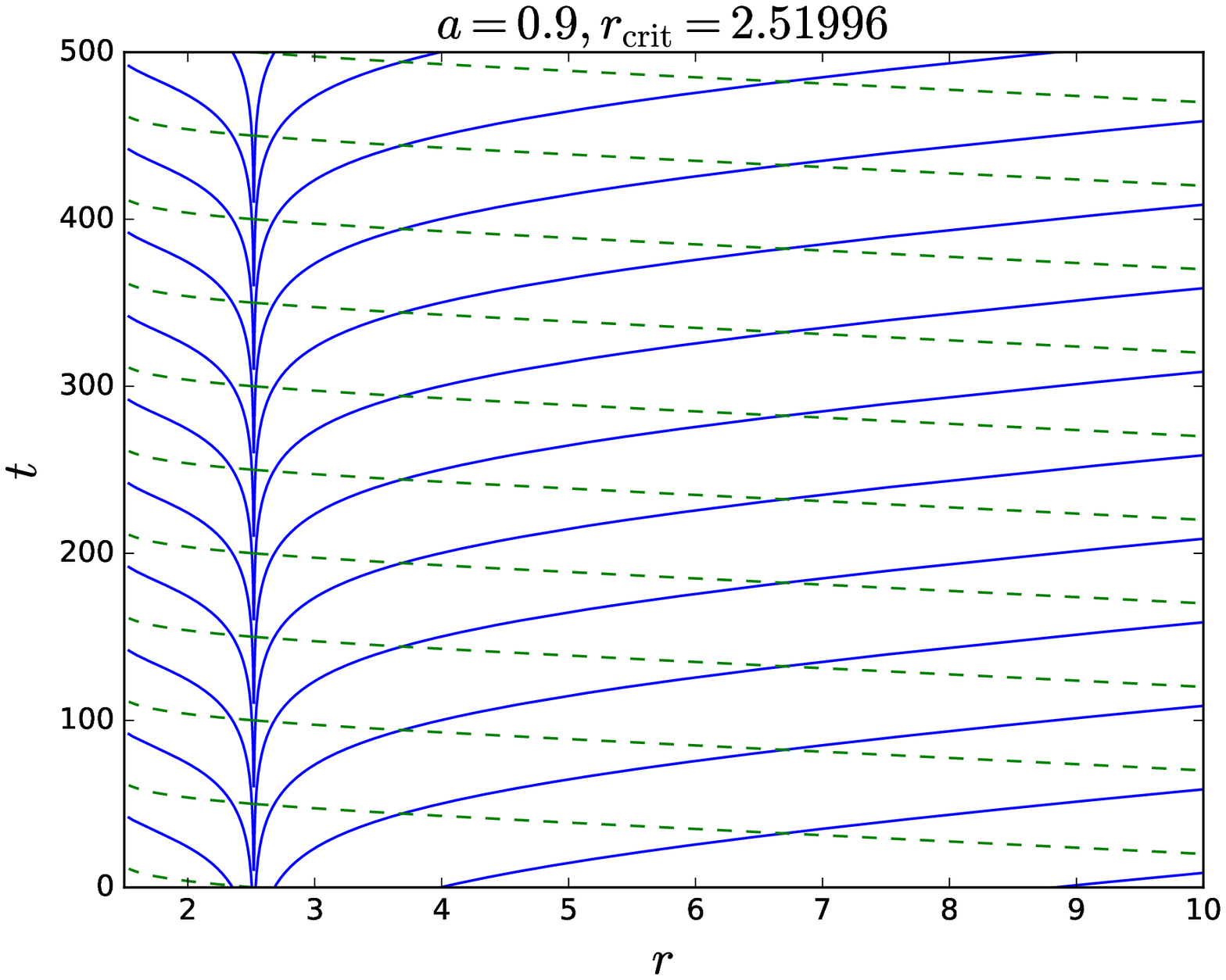}
	\end{tabular}
	\caption{ Causal structure of the acoustic spacetime for mono-transonic accretion. $ t_+(r) $ vs $ r $, i.e., $ z={\rm constant} $ lines are represented by the solid lines and $ t_-(r) $ vs $ r $, i.e., $ w={\rm constant} $ lines are represented by the dashed lines. $ t_\pm(r) $ are given by Eq.(\ref{t_pm}). The causal structures are plotted with $ [\xi_0,\gamma,\lambda_0]= [1.1,1.4,2.1] $ for $ a=-0.9,-0.5,0,0.5,0.9 $, row-wise from top to bottom. It could be noticed that the acoustic horizon where the $ t_+(r) $ lines diverges, coincides with the critical point $ r_{\rm crit} $. This further illustrates that the transonic surface of the accretion flow is indeed the acoustic horizon of the embedded acoustic spacetime.}\label{Fig:causal}
\end{figure*}
In Fig. \ref{Fig:causal}, we show the causal structure of the acoustic spacetime for mono-transonic accretion flow. The parameters $ [\xi_0,\gamma,\lambda_0] = [1.1,1.4,2.1]$ are same for all the plots while the black hole spins are $ a=-0.9,-0.5,0,0.5,0.9 $ row-wise from top to bottom. Solid lines represent $ t_+(r) $ vs $ r $, i.e., $ z={\rm constant} $ lines and the dotted lines represent the $ t_-(r) $ vs $ r $, i.e, $ w={\rm constant} $ lines. It is illustrated from the causal structures that the radius of the acoustic horizon, where $ t_+(r) $ diverges, are same as the critical points $ r_{\rm crit} $ for the given value of $ [\xi_0,\gamma,\lambda_0,a] $.
\subsection{Multi-transonic case}
For a given set of values of the parameters $ [\xi_0,\gamma,\lambda_0,a] $, the Eq. (\ref{critical_points})
can have more than one or more specifically three solutions for $ r>r_+ $. The corresponding flow in such case is said to be multi-critical flow as it allows multiple critical points $ r_{\rm in},r_{\rm mid},r_{\rm out} $ such that $ r_{\rm in}<r_{\rm mid}<r_{\rm out} $.  These critical points can be characterized by performing a critical point analysis. Such analysis shows that the inner and outer critical points $ r_{\rm in} $ and $ r_{\rm mid} $, respectively, are saddle type, whereas the middle critical point $ r_{\rm mid} $ is center type. Thus the accretion flow can pass only through the outer or inner critical points. when the accretion flow passes through both the outer and inner critical points, the accretion flow is called multi-transonic flow. Multi-critical flows are not necessarily multi-transonic flows. This could be understood as the following: suppose the flow starts its journey from large radial distance subsonically. At $ r=r_{\rm out} $, it makes a transition from subsonic state to supersonic state. Thus $ r_{\rm out} $ is basically the outer acoustic horizon. After the flow becomes supersonic it may encounter a shock formation which makes the flow subsonic from supersonic discontinuously, i.e., the dynamical variables such as the velocity, sound speed, density and pressure makes discontinuous jump. After it becomes subsonic due the shock formation, it again passes through the inner critical point and becomes supersonic from subsonic. Therefore in presence of shock formation, the flow can pass through both outer and inner critical points and hence the flow is multi-transonic. However, all the set parameters $ [\xi_0,\gamma,\lambda_0,a] $ which allow multiple critical points do not allow shock formation. In other words only a subset of the parameters allowing multiple critical points allow shock formation. This is best shown by plotting the parameter space. 

We have assumed a non-dissipative inviscid accretion flow. Therefore the flow has conserved specific energy and mass accretion rate. Thus the shock produced in such flow is assumed to be energy preserving Rankine Hugonoit type which satisfies the general relativistic Rankine Hugonoit conditions \cite{Eckart1940,Taub1948,Linchnerowicz1967,Thorne1973,Taub1978,Hacyan1982,abraham2006,Das2012}
\begin{equation}
\begin{aligned}
&[[\rho v^\mu \eta_\mu]]=[[\rho v^r]]=0\\
&[[T_{t\mu}\eta^\mu]] = [[(p+\varepsilon)v_{t}v^r]]=0\\
&[ [T_{\mu\nu}\eta^\mu\eta^\nu]]=[[(p+\varepsilon)(v^r)^2+pg^{rr}]]=0
\end{aligned}
\end{equation}
Where $ \eta_\mu=\delta^r_\mu $ is the normal to the surface of shock formation. $ [[f]] $ is defined as $ [[f] ]= f_+-f_- $ , where $ f_+ $ and $ f_- $ are values of $ f $ after and before the shock, respectively. First condition comes from the conservation of mass accretion rate and the last two conditions come from the energy-momentum conservation. These conditions are to be satisfied at the location of shock formation. In order to find out the  location of shock formation, it is convenient to construct a shock invariant quantity, which depends only on $ u_0,c_{s0} $ and $ \gamma $, using the conditions above. The first and second conditions are trivially satisfied owing to the constancy of the mass accretion rate and the specific energy. The first condition is basically $ (\Psi_0)_+ = (\Psi_0)_- \ $ and third condition is $ (T^{rr})_+=(T^{rr})_- $. Thus we can define a shock invariant quantity $ S_{\rm sh} = T^{rr}/\Psi_0 $ which also satisfies $ [[S_{\rm sh}]]=0 $ and is given by (see Appendix \ref{Sec:Sh-inv})
\begin{equation}\label{key}
S_{\rm sh} = \frac{(u_0^2(\gamma-c_{s0}^2)+c_{s0}^2)}{u_0\sqrt{1-u_0^2}(\gamma-1-c_{s0}^2)}.
\end{equation}

The procedure to find the location of shock formation is the following. Let us denote the values of $ S_{\rm sh} $ along the flow passing through outer critical point as $ S_{\rm sh}^{\rm out} $ and the same for the flow passing through inner critical point as $ S_{\rm sh}^{\rm in} $. At the location of shock formation $ r_{\rm sh} $, we have $ S_{\rm sh}^{\rm out}=S_{\rm sh}^{\rm in} $. Thus evaluating the $ S_{\rm sh}^{\rm out} $ and $ S_{\rm sh}^{\rm in} $ we find out $ r_{\rm sh} $ by noting the value of $ r $ for which $ S_{\rm sh}^{\rm out}=S_{\rm sh}^{\rm in} $. In general there are two such values of $ r_{\rm sh} $ such that one is between inner and middle critical points $ r_{\rm in} <r_{\rm sh1}<r_{\rm mid}$ and the other one is between middle and outer critical points $ r_{\rm mid} <r_{\rm sh2}<r_{\rm out}$. However, it has been shown in the literature that the shock formation at $ r_{\rm sh1} $ is unstable and that at $ r_{\rm sh2} $ is stable. Therefore only $ r_{\rm sh2} $ is the allowed location of shock formation and hence we shall refer to only this location as the location of shock formation, hereafter.

\begin{figure*}
	\centering
	\begin{tabular}{cc}
		%\resizebox{0.45\textwidth}{!}{\input{phase_portrait_a05}} & \resizebox{0.45\textwidth}{!}{\input{causal_w_shock_a05}} \\
		\includegraphics[scale = 0.4]{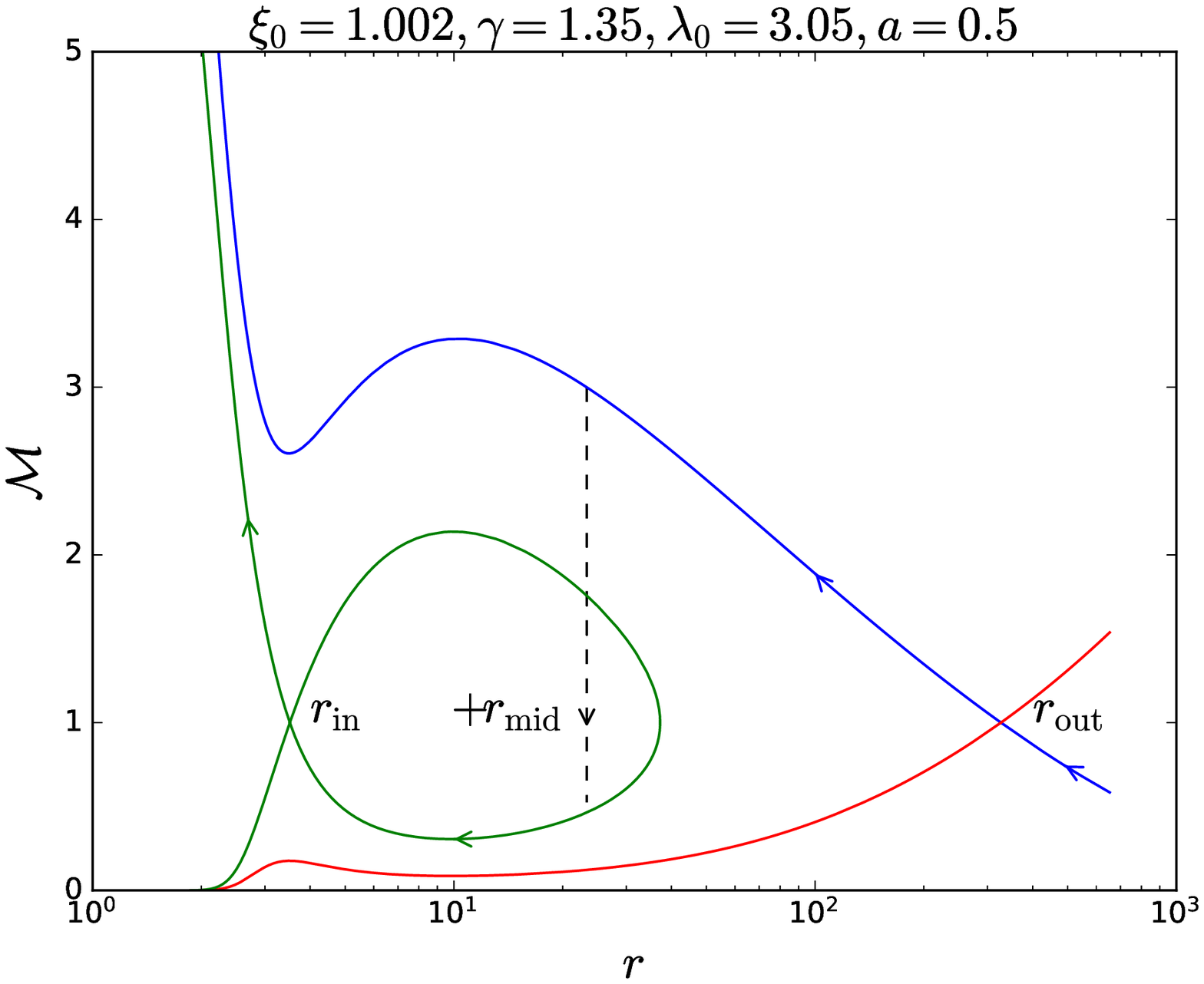} & \includegraphics[scale = 0.4]{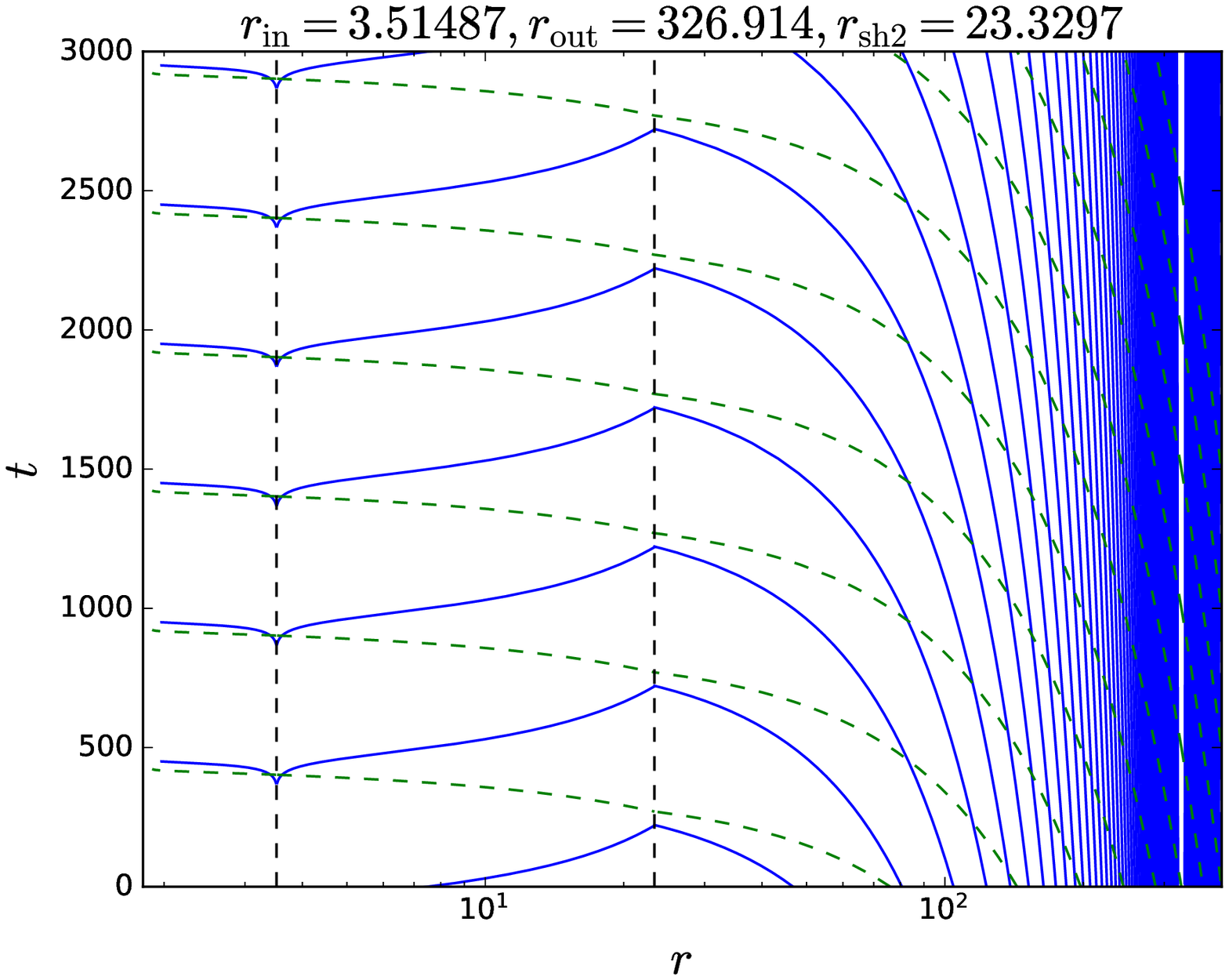}\\
		%\resizebox{0.45\textwidth}{!}{\input{phase_portrait_a055}} & 
		%\resizebox{0.45\textwidth}{!}{\input{causal_w_shock_a055}} \\
		\includegraphics[scale = 0.4]{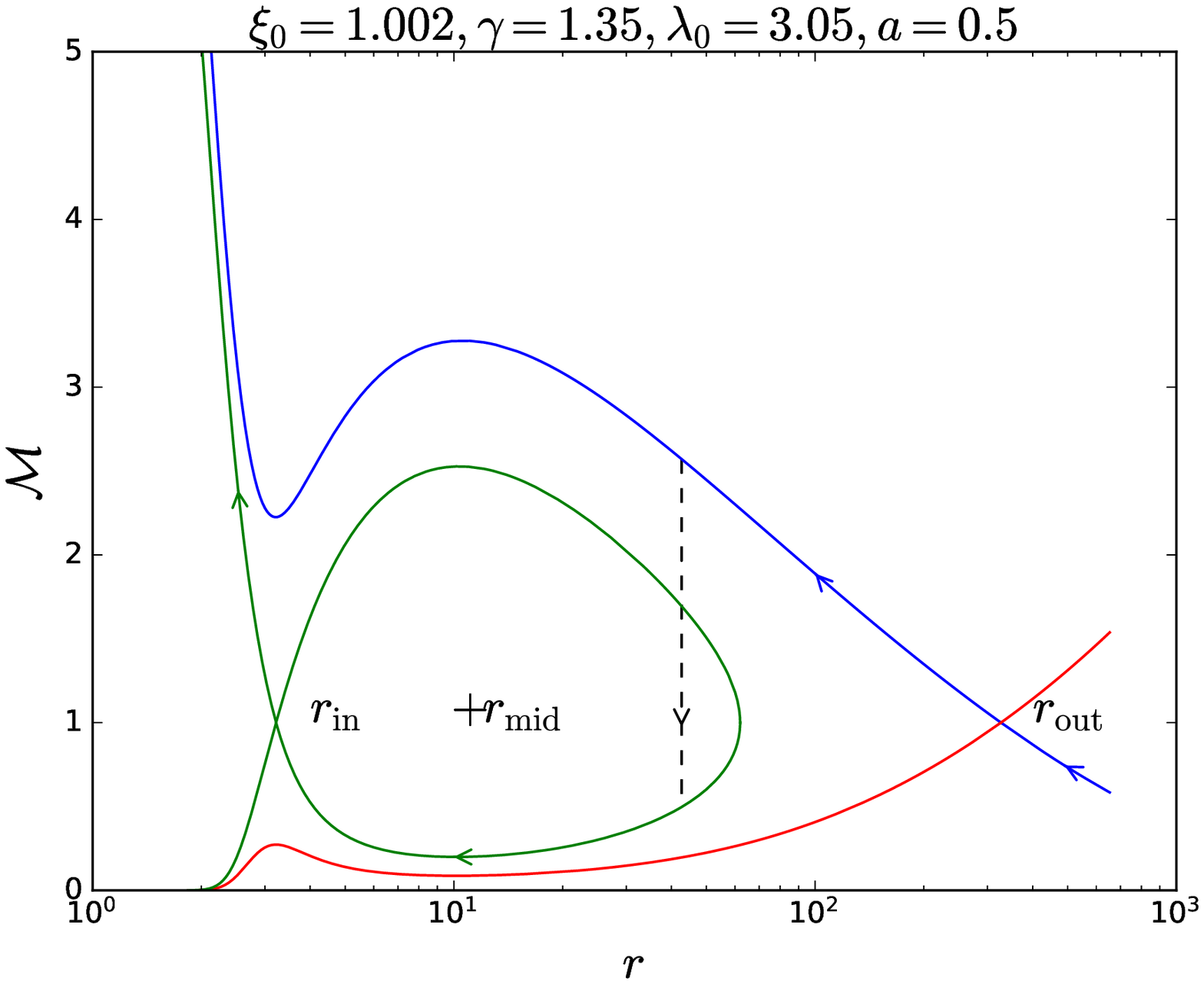} & \includegraphics[scale = 0.4]{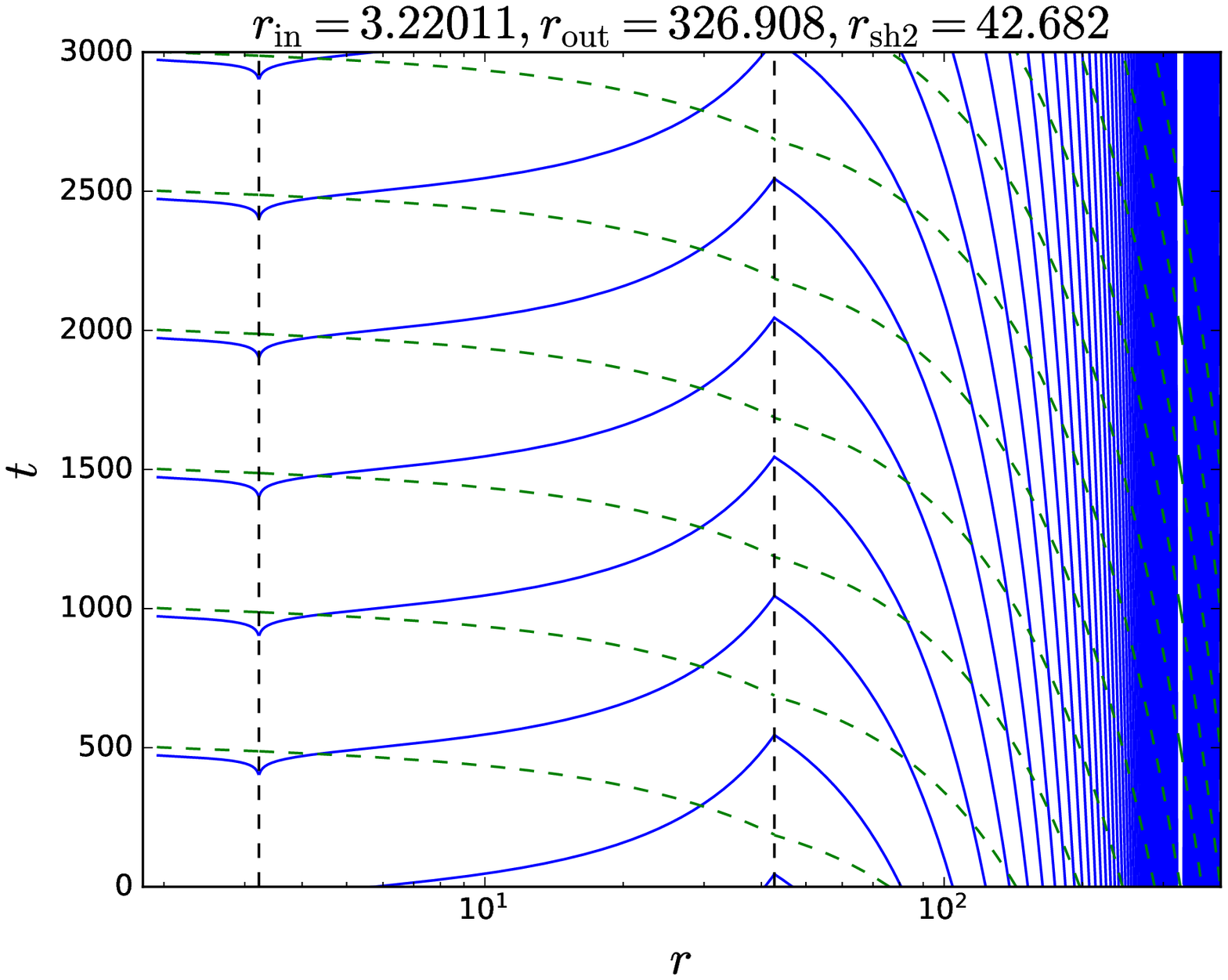} \\
		%\resizebox{0.45\textwidth}{!}{\input{phase_portrait_a06}} & 
		%\resizebox{0.45\textwidth}{!}{\input{causal_w_shock_a06}}\\
		\includegraphics[scale = 0.4]{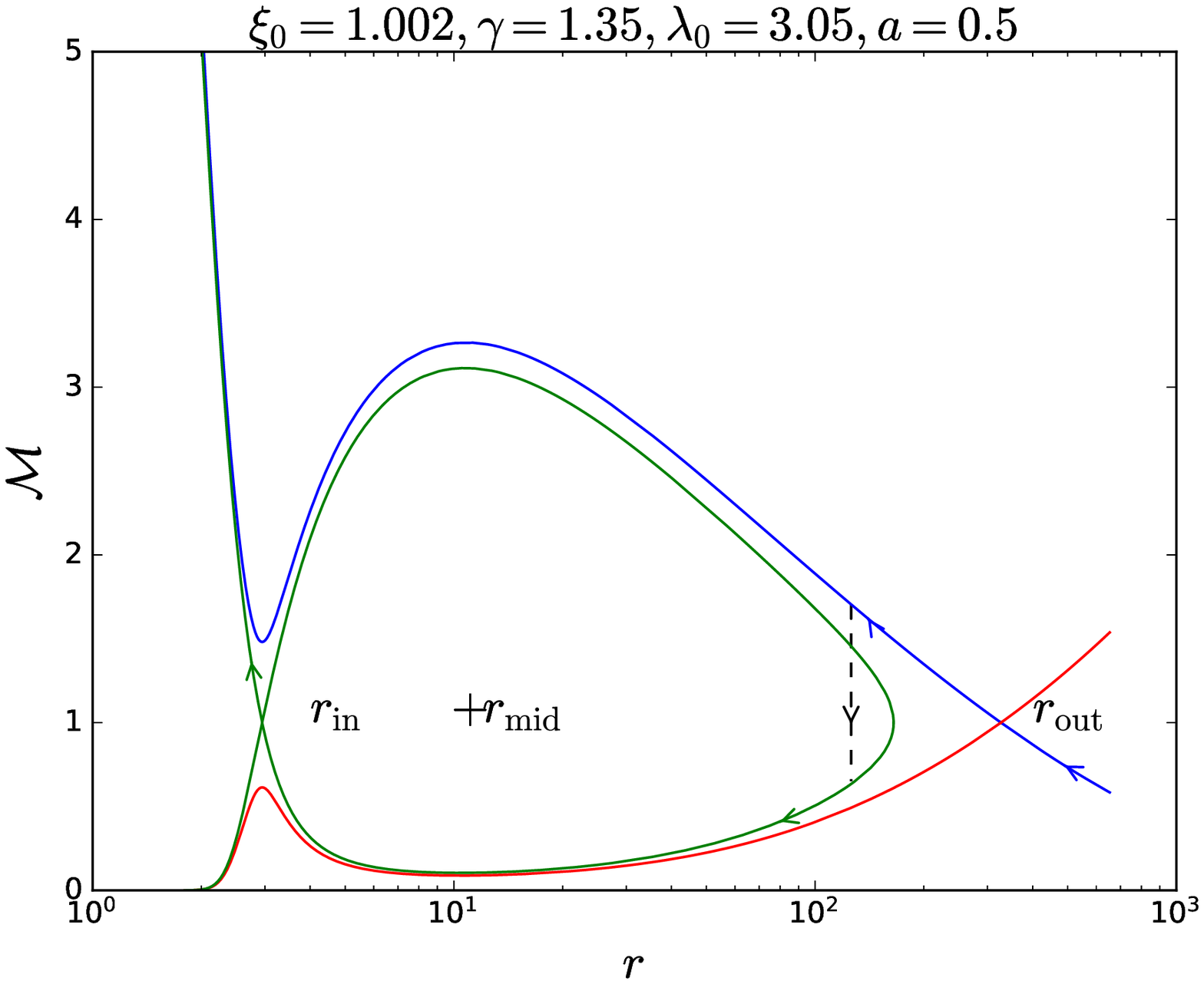} & \includegraphics[scale = 0.4]{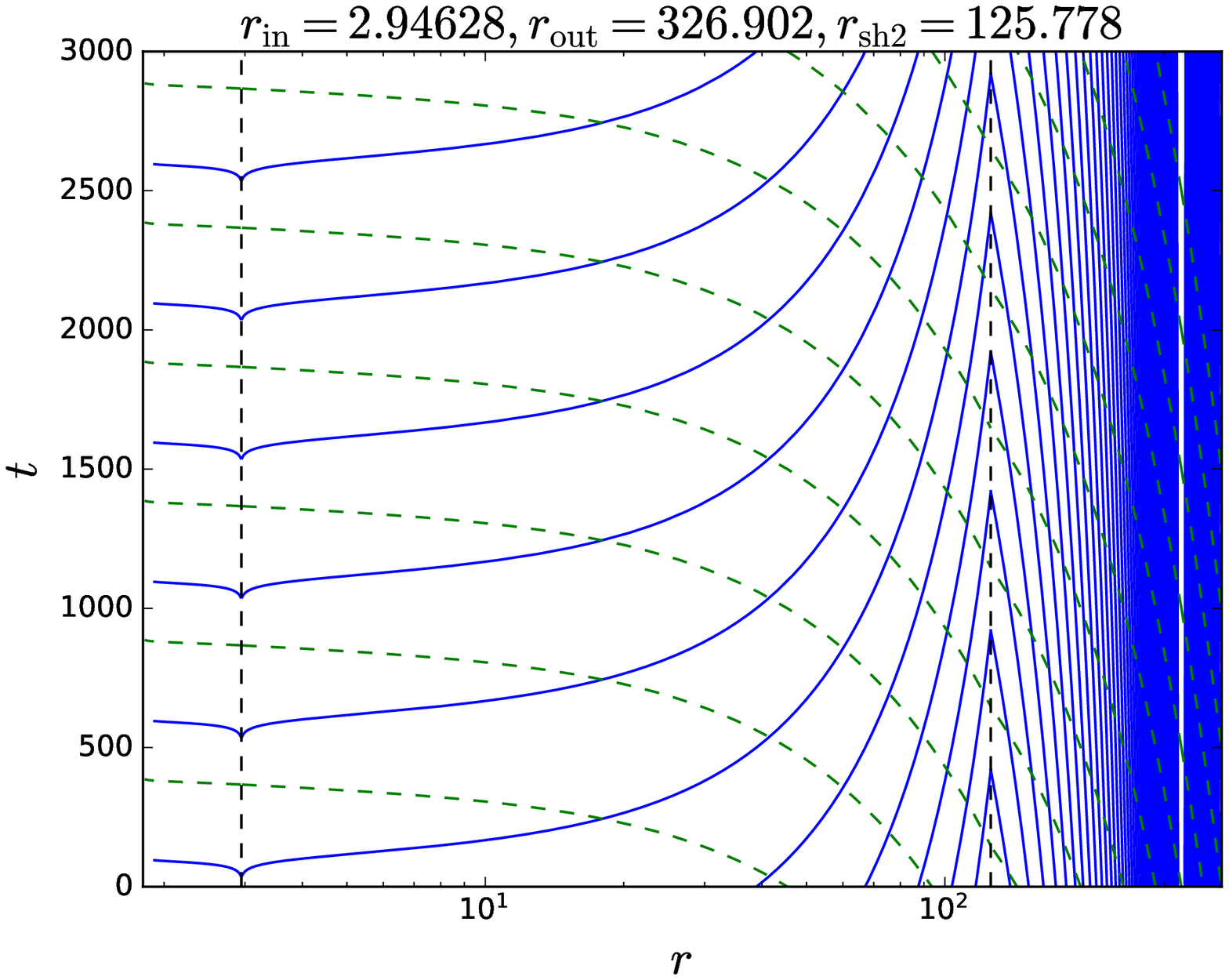} \\
	\end{tabular}
	\caption{ Mach number $ \mathcal{M} $ vs $ r $ plot (on the left) and the corresponding causal structure (on the right). The parameters $ [\xi_0 = 1.002,\gamma = 1.35, \lambda_0 = 3.05] $ are same where as the black hole spin is $ a = 0.5  $ (top panel), $ a = 0.55 $ (middle panel) and $ a = 0.6 $ (bottom panel). The  solid lines represents $ t_+(r) $ vs $ r $ lines and the dashed lines represents the $ t_-(r) $ vs $ r $ lines.}\label{Fig:multi-transonic}
\end{figure*}

In the left column of Fig. \ref{Fig:multi-transonic}, we show the phase portraits of the flow, i.e., the Mach number vs radial distance plots for three different values of the Kerr parameter $ a  = 0.5,0.55,0.6$, keeping $ [\xi_0,\gamma,\lambda_0] $ to be the same as $ [\xi_0 = 1.002,\gamma = 1.35,\lambda_0 = 3.05] $. These chosen values of the parameters $ [\xi_0,\gamma,\lambda_0,a] $ allow the flow to be multi-critical as well as multi-transonic by allowing shock formation. The shock transition of the flow has been denoted by a vertical dashed line in the phase portrait which implies that the shock formation at that location makes the flow to jump from supersonic state in the branch passing through the outer critical point to the subsonic state in the branch passing through the inner critical point. 

In the right column of the Fig. \ref{Fig:multi-transonic}, we show the causal structure corresponding the flow shown by the phase portrait in the left column in the particular row. In the causal structure plots, the vertical dashed line in the left is the location of the inner critical point and the vertical dashed line in the right is the location of shock formation. The outer critical point is located at the white line separating densely populated diverging $ t_+(r) $ lines. It is obvious from the causal structure that the inner and outer critical points are the inner and outer acoustic horizon of the acoustic spacetime. Also it could be noticed that for an observer in the region $ r_{\rm in} <r<r_{\rm sh2}$, the surface of shock formation would resemble a white hole horizon. Thus the shock formation can be regraded as the presence of an acoustic white hole.

\section{Acoustic surface gravity}
The Hawking temperature of an astrophysical black hole is given in terms of the surface gravity which can be derived by using the Killing vector which is null on the event horizon. Similarly, the analogue Hawking temperature $ T_{AH} $ may be given in terms of the acoustic surface gravity $ \kappa $ as $ T_{\rm AH} = \hbar\kappa/(2\pi K_B) $ in the units we are working with. $ K_B $ is the Boltzmann constant and $ \hbar=h/2\pi $, where $ h $ is the Planck constant. Suppose $ \chi^\mu $ is the Killing vector of the acoustic spacetime which is null on the acoustic horizon, i.e., $\chi^\mu\chi_\mu|_{\rm h}=G_{\mu\nu}\chi^\mu\chi^\nu|_{\rm h}=0$. Then the acoustic surface gravity is obtained by using the following relation (\cite{Poisson2004},\cite{Wald1984})
\begin{equation}\label{killing_eq}
\nabla_\alpha(-\chi^\mu\chi_\mu)=2\kappa\chi_\alpha
\end{equation}
The acoustic metric given by Eq. (\ref{acoustic_metric}) is independent of time $t$. Therefore we have the stationary Killing vector $\chi^\mu=\delta^\mu_t$ which is null on the horizon, i.e., $ G_{\mu\nu}\chi^\mu\chi^\nu|_{\rm h}= G_{tt}|_{\rm h} = 0 $. Now $\chi_\mu=G_{\mu\nu}\chi^\nu=G_{\mu\nu}\delta^\nu_t=G_{\mu t}$. Therefore from the $\alpha=r$ component of the Eq. (\ref{killing_eq}) the acoustic surface gravity is obtained to be
\begin{equation}
\kappa=\frac{1}{2G_{rt}}\partial_r(-G_{tt})|_{u_0^2=c_s^2}
\end{equation}
Using the expressions of $ G_{tt} $ and $ G_{rt} $ from Eq. (\ref{metric_elements_in_u_lambda}) provides
\begin{equation}\label{kappa1}
\kappa=\left|\kappa_0\left(\frac{du_0}{dr}-\frac{dc_{s0}}{dr}\right)\right|_{\rm h}
\end{equation}
where
\begin{equation}\label{key}
\kappa_0 = \frac{\sqrt{(g_{tt}g_{\phi\phi}+g_{\phi t}^2)(g_{\phi\phi}+2\lambda_0 g_{\phi t}-\lambda_0^2 g_{tt})}}{(1-c_{s0}^2)(g_{\phi\phi}+\lambda_0 g_{\phi t})\sqrt{g_{rr}}}
\end{equation}
and the subscript ``h", as mentioned earlier, denotes that the quantities have been evaluated at the acoustic horizon. On the equatorial plane ($\theta=\frac{\pi}{2}$) the metric elements are given by
\begin{equation}
g_{tt}=1-\frac{2}{r},\quad g_{\phi t}=-\frac{2a}{r},\quad g_{\phi\phi}=\frac{r^3+a^2r+2a^2}{r}
\end{equation}
Thus $\kappa_0$ can be further written as
\begin{equation}\label{kappa2}
\kappa_0=\frac{r\sqrt{(r^2-2r+a^2)(g_{\phi\phi}+2\lambda_0 g_{\phi t}-\lambda_0^2 g_{tt})}}{(1-c_{s0}^2)(r^3+a^2r+2a^2-2a\lambda_0)\sqrt{g_{rr}}}
\end{equation}
 The acoustic surface gravity is thus obtained as a function of the background metric elements and the stationary values of the accretion variables. The surface gravity depends explicitly on the black hole spin $a$. 

\section{Stability analysis}\label{sec:stability}
The wave equation describing the propagation of the mass accretion rate, as given by Eq. \ref{w_mass_final}, could be used to check whether the steady state accretion flow solutions are stable under linear perturbations. We discuss two different kind of solutions of the wave equation given by Eq. \ref{w_mass_final}. We follow the technique introduced by  \cite{Petterson1980} for this purpose.
Let us take the trial solution as 
\begin{equation}
 \Psi_1(r,t) = P_\omega (r) \exp[{i\omega t}] ,
 \end{equation} 
 using this trial solution in the wave equation $\partial_\mu(f^{\mu\nu}\partial_\nu \Psi_1)=0$, where $f^{\mu\nu}$ is  Eq. (\ref{f_mass}), provides
\begin{equation}\label{Stability_general}
 -\omega^2 f^{tt}P_\omega+i\omega[f^{tr}\partial_rP_\omega+\partial_r(f^{rt}P_\omega)]+\partial_r(f^{rr}\partial_rP_\omega)
 =0
 \end{equation} 

\subsection{Standing wave analysis}
In order to form a standing wave, the amplitude of the wave $ P_\omega(r) $ must vanish at two different radii $ r_1 $ and $ r_2 $ for all times, i.e., $ P_\omega(r_1)=0=P_\omega(r_2) $. The outer point $ r_2 $ could be located at the source at large distance from which accreting materials are coming. However, in order that the inner condition $ P_\omega(r_1) =0 $ is satisfied the accretor must have a physical surface. Also the solution must be continuous in the range $ r_1\leq r\leq r_2 $. If the accretor is a black hole then the accretion flow is necessarily supersonic at the event horizon (\cite{Frank1985,Liang1980}). Also there is no physical surface or mechanism to make the wave amplitude vanish at the horizon and hence in the case of a black hole, standing waves are not formed. Also if the accretion flow has a supersonic region then it is possible to develop shock at some radius and this would make the solution discontinuous. Therefore, in order that standing wave is formed the flow must be subsonic in the region $ r_1\leq r \leq r_2 $. For this reason we consider subsonic flow in the following. 

Multiplying the Eq. (\ref{Stability_general}) by $P_\omega(r)$ and integrating the resulting equation between $r_1$ and $r_2$ gives
\begin{equation}\label{multiplied_by_p}
\begin{aligned}
&\omega^2 \int_{r_1}^{r_2} P_\omega^2 f^{tt}dr-i\omega\int_{r_1}^{r_2}\partial_r[f^{tr}P_\omega^2]dr\\
&-\int_{r_1}^{r_2}[P_\omega \partial_r(f^{rr}\partial_r P_\omega)]dr=0
\end{aligned}
\end{equation}
Boundary conditions at $ r_1 $ and $ r_2 $ makes the middle term vanish and integrating the last term by parts, the Eq. (\ref{multiplied_by_p}) can be written as
\begin{equation}
\omega^2 \int_{r_1}^{r_2} P_\omega^2 f^{tt}dr+\int_{r_1}^{r_2} f^{rr}(\partial_r P_\omega)^2dr=0,
\end{equation}
which provides
\begin{equation}\label{omega_for_standing}
\omega^2 = -\frac{\int_{r_1}^{r_2} f^{rr}(\partial_r P_\omega)^2dr}{\int_{r_1}^{r_2} f^{tt} P_\omega^2 dr}
\end{equation}

From Eq. (\ref{f_mass}) $f^{tt}$ is given by
\begin{equation}
f^{tt} = \frac{g_{rr}v^r_0c_{s0}^2}{v^t_0 v_{t0}\tilde{\Lambda}}[-g^{tt}+(v^t_0)^2(1-\frac{1}{c_{s0}^2})]
\end{equation}
and $f^{rr}$ is given by
\begin{equation}
f^{rr} = \frac{g_{rr}v^r_0c_{s0}^2}{v^t_0 v_{t0}\tilde{\Lambda}}[g^{rr}+(v^r_0)^2(1-\frac{1}{c_{s0}^2})].
\end{equation}
as $ g^{tt}>0 $ and $ c_{s0}^2<1 $
\begin{equation}\label{key}
-g^{tt}+(v^t_0)^2(1-\frac{1}{c_{s0}^2})<0
\end{equation}
and using Eq. (\ref{v_in_CRF}) we have
\begin{equation}
\begin{aligned}
& g^{rr}+\frac{u_0^2}{g_{rr}(1-u_0^2)}(1-\frac{1}{c_{s0}^2}) = \frac{(1-u_0^2)+u_0^2(1-\frac{1}{c_{s0}^2})}{g_{rr}(1-u_0^2)}\\
& =\frac{\left(1-\frac{u_0^2}{c_{s0}^2}\right)}{g_{rr}(1-u_0^2)}>0
\end{aligned}
\end{equation}
Where we have used the fact that the accretion flow is subsonic $ u_0^2<c_{s0}^2 $. Hence $\omega^2>0$. Therefore $\omega$ has two real roots and the trial solution is oscillatory and the stationary accretion solution is stable.

\subsection{Traveling wave analysis} 
We consider traveling wave solution with the wavelength which is small compared to the characteristic radius of the accretor, which for the case of black hole could be taken as the radius of the event horizon. For such solutions, the frequency will be very large and hence the solution could be written as a power series of the following form 
\begin{equation}\label{series_solution}
P_\omega(r)= \exp \left[\sum_{n=-1}^{\infty} \frac{k_n(r)}{\omega^n} \right]
\end{equation}
Substituting the trail solution in Eq. (\ref{Stability_general}) enables us to find out leading order terms by equating the coefficients of individual power of $\omega$ to zero. Thus we get
\begin{eqnarray}
& \textrm{coefficient of } \omega^2:  f^{rr}(\partial_rk_{-1})^2  + 2i f^{tr}\partial_r k_{-1} -f^{tt} \nonumber \label{omega2}\\
&= 0\\ 
& \textrm{coefficient of } \omega: f^{rr}\left[\partial_r^2k_{-1}+2\partial_rk_{-1}k_0\right]+i[2f^{tr}\partial_r k_0 \nonumber \label{omega}\\ 
& +\partial_r f^{tr}]+\partial_rf^{rr}\partial_r k_{-1} = 0\\ 
& \textrm{coefficient of } \omega^0: f^{rr}[\partial_r^2 k_0 + 2\partial_r k_{-1}\partial_r k_1+(\partial_r k_0)^2]\nonumber \\
& +\partial_r f^{rr}\partial_r k_0+2i f^{tr}\partial_r k_1=0 \label{omega0}
\end{eqnarray}
Eq. (\ref{omega2}) gives 
\begin{equation}\label{k-1}
k_{-1}(r) = i \int \frac{-f^{tr}\pm \sqrt{(f^{tr})^2-f^{tt}f^{rr}}}{f^{rr}}dr
\end{equation}
using $k_{-1}(r)$ from Eq. (\ref{k-1}) in Eq. (\ref{omega}) gives 
\begin{equation}
\label{k0}
k_0(r) = -\frac{1}{2} \ln [\sqrt{(f^{tr})^2-f^{tt}f^{rr}}] + \textrm{constant}
\end{equation}
and using Eq. (\ref{k-1}) and (\ref{k0}) in Eq. (\ref{omega0}) gives 
\begin{equation}
\label{k1}
k_1(r) =\pm \frac{i}{2} \int \frac{\partial_r(f^{rr}\partial_r k_0)+f^{rr}(\partial_r k_0)^2}{\sqrt{(f^{tr})^2-f^{tt}f^{rr}}}dr 
\end{equation}. Now
\begin{equation}
\textrm{det} f^{\mu\nu} = f^{tt}f^{rr}-(f^{rt})^2 =\left( \frac{g_{rr}v_0^r c_s^2}{v^t_0 v_{t0}\tilde{\Lambda}}\right)^2\mathcal{F}
\end{equation}
where $v_{t0}$ is the stationary value of $v_t$ given by Eq. (\ref{v_t}) and $v^r_0$ and $v^t_0$ are stationary values of $v^r$ and $v^t$ given by Eq. (\ref{v_in_CRF}) and (\ref{vt_0}) respectively and
\begin{equation}\label{stability_F}
\mathcal{F}=[-g^{tt}g^{rr}+(1-\frac{1}{c_s^2})(-g^{tt}(v^r_0)^2+g^{rr}(v^t_0)^2)].
\end{equation}
In terms of $\lambda_0,u_0$ and the background metric elements  $\mathcal{F}$ can be written as
\begin{equation}
\begin{aligned}
&\mathcal{F} = -\frac{g_{\phi\phi}}{g_{rr}(g_{\phi\phi}g_{tt}+g_{\phi t}^2)}\times \\
& \left[1+\frac{(1-c_s^2)}{c_s^2(1-u_0^2)}\left(\frac{(1+\lambda_0 \frac{g_{\phi t}}{g_{\phi \phi}})^2}{(1+2\lambda_0 \frac{g_{\phi t}}{g_{\phi\phi}}-\lambda_0^2 \frac{g_{tt}}{g_{\phi\phi}})} -u_0^2\right) \right] < 0
\end{aligned}
\end{equation}
Thus $k_{-1}(r)$ and $k_1(r)$ are purely imaginary and the leading contribution to the amplitude of the wave comes from $k_0(r)$. 

In order that the trial solution does not diverge and is stable, the power series in Eq. (\ref{series_solution}) must converge, i.e., we have to show $ |k_n/\omega_n| \gg |k_{n+1}/\omega_{n+1}|  $. As the frequency is very large $ \omega\gg 1 $, the contributions from higher order terms are very small. Thus it should suffice to show that $ |\omega k_{-1}|\gg |k_0|\gg |k_1/\omega| $. $ k_{-1},k_0,k_1 $ are complicated functions of the accretion variables and thus it is not possible to have an analytic form. However, we can find the spatial dependence at large distance $ r\to\infty $ where the spacetime is effectively Newtonian. From the constancy of the mass accretion rate we have $ v^r \propto 1/(\rho r^2) $. At the asymptotic limit $ \rho $ approaches its constant ambient value $ \rho_\infty $ and hence at $ r\to \infty $, $ v^r_\infty \propto 1/r^2 $. Similarly the sound speed has its ambient value $ c_{s0\infty} $. $ v^{t}_0\sim 1 $ and $ v_{t0}\sim -1 $. Also $ \tilde{\Lambda}_\infty \propto (v^r_0)^2 $. Thus 
In this asymptotic limit we have
\begin{equation}\label{key}
\begin{aligned}
& f^{tt}\sim r^2,\quad f^{rr}\sim r^2,\quad f^{tr}\sim r^0, (f^{rt})^2-f^{tt}f^{rr}\sim r^4
\end{aligned}
\end{equation} 
which gives $ k_{-1} \sim r$, $ k_0\sim \ln r $ and $ k_1\sim 1/r $. Therefore, the sequence converges in the leading order even at large $ r $.
Considering the first three terms in the expansion in Eq. (\ref{series_solution}) the amplitude of the wave can be approximated as
\begin{equation}
\label{wave_amplitude}
 |\Psi_1|\approx \left[\left(\frac{v^t_0v_{t0}\tilde{\Lambda}}{g_{rr}v^r_0 c_s^2}\right)^2 \frac{1}{-\mathcal{F}}\right]^{\frac{1}{4}}
\end{equation}

\section{Concluding remarks}
In this work, we demonstrate that the emergence of acoustic spacetime as an analogue system is a natural outcome of the linear stability analysis of the relativistic black hole accretion. It is interesting to investigate whether, in general, the emergence of gravity like phenomena is a consequence of linear perturbation analysis only, or any complex nonlinear perturbation (of any order) of fluid may lead to the emergent gravity phenomena. In other words, it is important to know how universal the analogue gravity phenomena is -- whether black hole like spacetime can be generated by only one means (linear perturbation) or any kind of perturbation of general nature would lead to the construction of an analogue system. In another work (\cite{Nandan2018}), we have started explaining this for standard Newtonian fluid flow. In our future work, we would like to explore the possibility of obtaining (or not) an acoustic spacetime through the process of higher order perturbation analysis of relativistic astrophysical accretion. It is to be noted that the correspondence between the analogue system and the accretion astrophysics can be established through the process of linear stability analysis of stationary integral accretion solutions. That means, only the steady state accretion has been considered. The body of literature in accretion astrophysics is huge and diverse, and hence there are several excellent works that exist in literature where complete time-dependent numerical simulation has been performed to study non-steady flow of hydrodynamic fluid including various kind of time variabilities (\cite{Hawley1984a,Hawley1984b,Kheyfets1990,Hawley1991,Yokosawa1995,Igumenshchev1996,Igumenshchev1997,Nobuta1999,Molteni1999,Stone1999,Caunt2001,DeVilliers2002,Proga2003,Gerardi2005,Moscibrodzka2008,Nagakura2008,Nagakura2009,Janiuk2009,Bambi2010a,Bambi2010b,Barai2011,Barai2012,Sukova2015MNRAS,Zhu2015,Narayan2016,Moscibrodzka2016,Sukova2017,Sadowki2017,Mach2018,Karkowski2018,Inayoshi2018,Fragile2018}). Also in our present work, we limit our stability analysis procedure within a purely analytical framework and did not opt for any numerical studies in this aspect. There are, however, a number of works exist in the literature (for some recent works, see \cite{Penner2011,Lora-Clavijo2013,Gracia-Linares2015,Gonzalez2018}) which studies, fully numerically, the stability analysis of spherically or axially symmetric black hole accretion in two or three dimensions. We, however, did not concentrate on such approach since our main motivation was to explore how the emergent gravity phenomena can be observed through the stability analysis of steady-state solutions of hydrodynamic accretion.

In the present work, we have explicitly performed the perturbation analysis to make correspondence between the analogue gravity and the accretion astrophysics around black holes. Various properties of the corresponding analogue spacetime, however, can be studied by examining the stationary solutions as well, both for matter flow in spherically symmetric as well as for axially symmetric accretion (\cite{Das2004,Dasgupta2005,Das2007,abraham2006,Pu2012,Bilic2014,Tarafdar2015,Tarafdar2018}). 

In theoretical physics, one of the main objectives to study the analogue gravity phenomenon is to understand the Hawking like effects -- the emission of phonons from the close vicinity of the acoustic horizon which is considered to be analogous to the usual Hawking radiation emanating out from the standard gravitational black holes.

Even though the detailed analysis of quantum Hawking like effects may not be possible in a purely classical analogue system, the study of the acoustic surface gravity may have significant importance in such systems. The acoustic surface gravity itself is an important entity to study as it may help to understand the flow structure as well as the acoustic spacetime. Therefore, the acoustic surface gravity may be studied independently without studying the analogue Hawking radiation like phenomena characterized by the analogue Hawking temperature which may be too small to be detected experimentally in such system. The acoustic surface gravity plays an important role to study the non-negligible effects associated with the analogue Hawking effects which could be examined through the modified dispersion relations. Such studied has been performed in purely analytical work as well as experimental setup (\cite{Rousseaux2008,Rousseaux2010,Jannes2011,WeinfurtnerPRL2011,Leonhardt2012,Robertson2012}). 

The deviation of the Hawking like effect in the dispersive medium depends very sensitively on the gradient of dynamical velocity as well as that of the sound speed. In most of the above-mentioned studies, the sound speed is taken to constant or in other words the flow is taken to be isothermal. Also, the velocity gradient is estimated by prescribing a particular velocity profile using certain assumptions. On the other hand in our current work, the values of the space gradient of both the dynamical flow velocity and the speed of sound have been computed very accurately. Thus it is obvious that the non-universal feature of the Hawking like effect could be further modified by studying the black hole accretion system as an analogue gravity system. Therefore, it is obvious that though the accreting black hole system may not provide any direct signature of the Hawking like effect, it can still be considered as a very important as well unique theoretical construct to study analogue gravity phenomena.

Lastly, one may argue that the analogue Hawking temperature may be significant in case of accretion around a primordial black hole. However, the accretion process on primordial black holes itself is an area which is not completely understood. To study accretion in such system one has to first construct a self-consistent model of accretion onto such primordial black holes. Such study is clearly beyond the scope of the present work and hence we concentrate only on large astrophysical black holes.

\section{Acknowledgments}
MAS sincerely expresses his deep gratitude to Sourav Bhattacharaya for his very useful help and guidance. TKD acknowledges the support from the Physics and Applied Mathematics Unit, Indian Statistical Institute, Kolkata, India, in the form of a long-term visiting scientist (one-year sabbatical visitor). The authors thank the anonymous referees for usefull comments and suggestions. 

\appendix
\section{Accretion flow equations}\label{Sec:flow-eq}
To derive the expression for the gradient of advective velocity $ du_0/dr $ and the gradient of the sound speed $ dc_{s0}/dr $
we use the expressions for the two conserved quantities of the flow. The mass accretion rate $ \Psi_0 $ in terms of $ u_0 $ is given by
\begin{equation}\label{mdot}
\Psi_0 = 4\pi H_0 r^2 \rho_0 \frac{u_0}{\sqrt{g_{rr}(1-u_0^2)}}
\end{equation}
and the relativistic Bernoulli's constant is given by

\begin{equation}\label{xi-ad}
\xi_0 = h_0\sqrt{\frac{g_{tt}g_{\phi\phi}+g_{\phi t}^2}{(g_{\phi\phi}+2\lambda_0 g_{\phi t}-\lambda_0^2 g_{tt})(1-u^2)}}.
\end{equation}
For adiabatic flow with conserved specfic entropy, in other words an isentropic flow, the enthalpy is given by $ dh = dp/\rho $ which when used in the definition of enthalpy given $ h = (p+\varepsilon)/\rho $ gives $ h = d\varepsilon/d\rho $. The energy density $ \varepsilon $ includes rest-mass energy $ \rho $ and an internal energy equal to $ p/(\gamma-1) $. Thus $ \varepsilon = \rho + p/(\gamma-1) $. For polytropic equation of state $ p = k\rho^\gamma $, the enthalpy is therefore given by
\begin{equation}\label{enthalpy}
h_0 = \frac{\gamma-1}{\gamma-(1+c_{s0}^2)}
\end{equation}
To obtain an equation for the gradient of the sound speed one defines a new quantity $ \dot{\Xi} $ via the following transformation
\begin{equation}
\dot{\Xi} = \Psi_0(\gamma k)^{\frac{1}{\gamma-1}}
\end{equation}
$ k $ is a measure of the specific entropy of the accreting matter as the entropy per particle $ \sigma $ is related to $ k $ as
\begin{equation}
\sigma = \frac{1}{\gamma-1}\log k+\frac{\gamma}{\gamma-1}+{\rm constant}.
\end{equation}
Thus $ \dot{\Xi} $ represents the total inward entropy flux and could be labelled as the stationary entropy accretion rate. Expressing $ \rho $ in terms of $ k,\gamma,h,c_{s0}^2 $, the entropy accretion rate could be written as
\begin{equation}
\dot{\Xi} = 4\pi H_0 \frac{u_0}{\sqrt{g_{rr}(1-u_0^2)}}r^2 \left(\frac{(\gamma-1)c_{s0}^2}{\gamma-(1+c_{s0}^2)}\right)^{\frac{1}{\gamma-1}}
\end{equation}
Taking the logarithmic derivative of the above equation with respect to $ r $, the gradient of the sound speed could be written as
\begin{equation}\label{dcdr}
\frac{dc_{s0}}{dr} = -\frac{c_{s0}(\gamma-(1+c_{s0}^2))}{2}\left[\frac{1}{u_0(1-u_0^2)}\frac{du_0}{dr}+\frac{1}{r}+\frac{1}{2}\frac{\Delta'}{\Delta}\right]
\end{equation}
where $ \Delta  = r^2-2r+a^2$ as given by Eq. (\ref{A-B-C}) and the $\Delta'$ denotes the first derivative of $\Delta$ with respect to $r$. The gradient of the advective velocity could be found by taking logarithmic derivative of eq. (\ref{mdot}) and eq. (\ref{xi-ad}) (substituting $ dh/h_0 = c_{s0}^2 d\rho/\rho_0 $) and eliminating $ d\rho/\rho_0 $, which gives 
\begin{equation}\label{dudr}
\frac{du_0}{dr} = \frac{u_0(1-u_0^2)\left[c_{s0}^2\left(\frac{2}{r}+\frac{\Delta'}{\Delta}\right)-\frac{\Delta'}{\Delta}+\frac{B'}{B}\right]}{2(u_0^2-c_{s0}^2)} = \frac{N}{D}
\end{equation}
where $ B =  (g_{\phi\phi}+2\lambda_0 g_{\phi t}-\lambda_0^2 g_{tt})$ and $B'$ is the first derivative of $B$ with respect to $r$. The critical points of the flow are obtained by equating $ D= N =0 $. $ D = 0$ gives the location of critical points to at $ u_0^2|_{\rm crit}=c_{s0}^2|_{\rm crit} $. and $ N=0 $ gives 
\begin{equation}
\left.u_0^2\right|_{r=r_{\rm crit}} = \left.c_{s0}^2\right|_{r=r_{\rm crit}} =\left. \frac{\frac{\Delta'}{\Delta}-\frac{B'}{B}}{\frac{2}{r}+\frac{\Delta'}{\Delta}}\right|_{r = r_{\rm crit}}
\end{equation}
Using the above condition we can substitute $ u_0^2 $ and $ c_{s0}^2 $ in eq. (\ref{xi-ad}) at the critical points which provides
\begin{equation}\label{critical_points}
\xi_0 = \left.\frac{1}{1-\frac{1}{\gamma-1}\frac{\frac{\Delta'}{\Delta}-\frac{B'}{B}}{\frac{2}{r}+\frac{\Delta'}{\Delta}}}\sqrt{\frac{2\Delta+r\Delta'}{2B+rB'}}\right|_{r=r_{\rm crit}}
\end{equation}
Thus, for a given value of $ \xi_0 $ which is a constant along the flow and that of $ \gamma,\lambda_0 $ and $a $, the above equation could be solved for $ r_{\rm crit} $ numerically and the critical points could be found. To find the value of the gradient of the advective velocity at the critical points,  we use L'Hospital rule which gives
\begin{equation}
\left. \frac{du_0}{dr}\right|_{\rm crit} = \frac{-\beta\pm\sqrt{\beta^2-4\alpha\Gamma}}{2\alpha}
\end{equation}
where
\begin{equation}
\begin{aligned}
& \alpha = \left.1+\gamma-3c_{s0}^2\right|_{r=r_{\rm crit}} \\
& \beta = \left.2c_{s0}(1-c_{s0}^2)(\gamma-(1+c_{s0}^2))\left(\frac{1}{r}+\frac{\Delta'}{2\Delta}\right)\right|_{r=r_{\rm crit}}\\
& \Gamma = c_{s0}^2(1-c_{s0}^2)^2\left[(\gamma-(1+c_{s0}^2))\left(\frac{1}{r}+\frac{\Delta}{2\Delta'}\right)^2-\Gamma^1\right]_{r = r_{\rm crit}}\\
& \Gamma^1 = \frac{1-c_{s0}^2}{2c_{s0}^2}\left(\frac{\Delta'^2}{\Delta^2}-\frac{\Delta''}{\Delta}\right)-\frac{1}{2c_{s0}^2}\left(\frac{B'^2}{B^2}-\frac{B''}{B}\right)-\frac{1}{r^2}
\end{aligned}
\end{equation}
$\Delta''$ and $B''$ are the second derivatives of $\Delta$ and $B$ with resepect to $r$, respectively. For a given set of paramters $ [\xi_0,\gamma,\lambda_0,a] $, we can now solve eq. (\ref{dudr}) and (\ref{dcdr}) simultaneously to obtain the Mach number as a function of the radial coordinate $ r $. Depending on the values of the parameters $ [\xi_0,\gamma,\lambda_0,a] $, the phase portrait may contain one or more critical points. 

\section{Shock invariant quantity}\label{Sec:Sh-inv}
$ h_0 $ is given by equation Eq. (\ref{enthalpy}). $ c_{s0}^2=(1/h_0)dp/d\rho = (1/h_0)k\gamma \rho^{\gamma-1}$, which gives $ \rho_0 $ (and hence also $ p $ and $ \varepsilon $) in terms of $ k,\gamma $ and $ c_{s0} $. Thus
\begin{equation}
\begin{aligned}
& \rho = k^{-\frac{1}{\gamma-1}}\left[\frac{(\gamma-1)c_{s0}^2}{\gamma(\gamma-1-c_{s0}^2)}\right]^{\frac{1}{\gamma-1}}\\
& p = k^{-\frac{1}{\gamma-1}}\left[\frac{(\gamma-1)c_{s0}^2}{\gamma(\gamma-1-c_{s0}^2)}\right]^{\frac{\gamma}{\gamma-1}}\\
& \varepsilon = k^{-\frac{1}{\gamma-1}}\left[\frac{(\gamma-1)c_{s0}^2}{\gamma(\gamma-1-c_{s0}^2)}\right]^{\frac{1}{\gamma-1}}\left(1+\frac{c_{s0}^2}{\gamma(\gamma-1-c_{s0}^2)}\right)
\end{aligned}
\end{equation}
Now $ \Psi_0 = {\rm constant}\times r^2 \rho v^r_0 $ and $ T^{rr}=(p+\varepsilon)(v^r_0)^2+pg^{rr} $, where $ v^r_0 = u_0/\sqrt{g_{rr}(1-u_0^2)}  $. Therefore the shock-invariant quantity $ S_{\rm sh} = T^{rr}/\Psi_0 $ becomes
\begin{equation}\label{key}
S_{\rm sh} = \frac{(u_0^2(\gamma-c_{s0}^2)+c_{s0}^2)}{u_0\sqrt{1-u_0^2}(\gamma-1-c_{s0}^2)}
\end{equation}
where we have remove any over all factor of $ r $ as shock invariant quantity is to be evaluated at constant $ r=r_{\rm sh} $.

\bibliography{reference_arif} 

\end{document}